\documentclass[11pt,a4paper]{article}
\usepackage{jheppub_kim}

\usepackage{pdflscape}
\usepackage{amsmath}
\usepackage{amssymb}
\usepackage{dcolumn}
\usepackage{bm}
\usepackage{color}
\usepackage{epsfig}
\usepackage{amsfonts}
\usepackage{graphicx}
\usepackage{subfigure}
\usepackage{dcolumn}
\newcommand{\tcb}{\textcolor{blue}}
\newcommand{\tcr}{\textcolor{red}}

\begin{document}

\title{Thermal fluctuations of (non)linearly charged BTZ black hole in massive gravity}
\author[a]{Behnam Pourhassan,}
\author[c,b]{Seyed Hossein Hendi,}
\author[d,a]{Sudhaker Upadhyay\footnote{Corresponding author}\footnote{Visiting Associate, Inter-University Centre for Astronomy and Astrophysics
(IUCAA) Pune, Maharashtra-411007},}
\author[e]{\.{I}zzet Sakall{\i}}
\author[f,g,h]{Emmanuel N. Saridakis,}

\affiliation[a] {School of Physics, Damghan University, Damghan,
3671641167, Iran}

\affiliation[b] {Canadian Quantum Research
Center 204-3002 32 Ave Vernon, BC V1T 2L7 Canada}

\affiliation[c] {Physics Department and Biruni Observatory,
College of Sciences, Shiraz University, Shiraz 71454, Iran  }

\affiliation[d] {Department of Physics, K. L. S. College, Magadh
University, Nawada-805110,  India}

\affiliation[e] {Department of Physics, Eastern Mediterranean University, 99628 Famagusta,
Cyprus}

\affiliation[f]{National Observatory of Athens, Lofos Nymfon, 11852 Athens, Greece.}
\affiliation[g]{Deep Space Exploration Laboratory/School of Physical Sciences, University of Science and
Technology of China, Hefei, Anhui 230026, China}

\affiliation[h]{Departamento de Matem\'{a}ticas, Universidad Cat\'{o}lica del Norte, Avda. Angamos 0610,
Casilla 1280 Antofagasta, Chile}

\emailAdd{b.pourhassan@du.ac.ir} \emailAdd{hendi@shirazu.ac.ir}
\emailAdd{sudhakerupadhyay@gmail.com} \emailAdd{izzet.sakalli@emu.edu.tr}
\emailAdd{msaridak@noa.gr}

\abstract{We consider a charged BTZ black hole in
asymptotically AdS space-time of massive gravity to study the
effect of the thermal fluctuations on the black hole
thermodynamics. We consider the Einstein-Born-Infeld solution and
investigate critical points and stability. We also compare the
results with the case of Einstein-Maxwell solutions. Besides, we
find that thermal fluctuations, which appear as a logarithmic term
in the entropy, affect the stability of the black hole and change the
phase transition point. Moreover, we study the geometrical
thermodynamics and find that the behaviour of the linear Maxwell
solution is the same as the nonlinear one.}


\maketitle

\section{Introduction}
The quest to understand the fundamental laws of nature has been one of the most significant scientific endeavours in human history.
One of the most exciting and challenging areas of study in modern physics is the investigation of the interplay between gravity
and other fundamental forces. The study of (2+1) Einstein gravity, which is the three-dimensional analogue of the more familiar
(3+1) Einstein's gravity has been of great interest to physicists for many years. One of the reasons for this is that (2+1) Einstein gravity
is more tractable than its $(3+1)$ counterpart and this makes it a useful testbed for exploring a wide range of ideas and concepts
in gravitational physics. To decrease the algebraic complexity of calculations, one may consider $(2+1)-$dimensions instead of four-dimensional counterpart. For example, AdS3/CFT2 correspondence is one of the best-understood concepts of holography in three-dimensional gravitating systems
\cite{BasuPRD2016}. Besides, the holographic description of non-relativistic strongly coupled systems,
such as what happens in condensed matter physics, has been well investigated in three-dimensional spacetime
\cite{JaniszewskiJHEP2013,GriffinPRL2013}. Furthermore, studying the three-dimensional solutions helps us to understand a profound insight
in black hole physics, quantum gravity viewpoint and also its relations to string theory
\cite{Carlip1995, Ashtekar2002,Sarkar2006,Witten1998,Carlip2005}.\\
The addition of massive gravity to (2+1) Einstein's gravity is particularly interesting because
it introduces new degrees of freedom into the theory, which can have significant implications for the behaviour of the gravitational field.
Massive gravity is a relatively new field of study, and it involves introducing a mass term into the Einstein equations,
which modifies the behaviour of the gravitational field at large distances. This modification can have important implications for how gravitational waves propagate and interact with matter, and it is an area of active research.
In addition to new massive gravity \cite{Newmassive} theory, which is formulated just in three dimensions,
there is another ghost-free massive gravity, proposed by de Rham, Gabadadze and Tolley (dRGT) \cite{deRham:2010kj}, that can be organized in arbitrary dimensions \cite{Kanzi:2020cyv}.
A subclass of dRGT theory was introduced by Vegh \cite{Vegh} with the applications of gauge/gravity duality.
Using this theory of massive gravity, Vegh showed that the massive graviton may behave like a lattice and exhibit a Drude peak \cite{Chen:2017dsy}. In addition, it was shown that for an arbitrary singular metric, this theory of massive gravity is ghost-free and stable \cite{Zhang2016}.
Besides, the effects of the mass of graviton in Vegh's massive gravity model on the maximal tunnelling current and the
the coherence length of junction in the metal/superconductor phase transition is investigated \cite{Hu2016}.
Moreover, an interesting investigation of the phase transition of holographic entanglement entropy in this massive gravity
has been reported in \cite{Zeng2016}. Therefore, one may be motivated to study black hole solutions in the presence of Vegh's
massive gravity in three and higher dimensions. Non-linear electrodynamics, on the other hand, is concerned with the study of electromagnetic fields in the presence of strong fields \cite{ras,ras1,ras2,ras3,ras4,ras5,ras6,ras7,ras8}. { 
In the realm of theoretical physics, the Born–Infeld model \cite{Born:1933pep,
Born:1934gh}, also known as the Dirac–Born–Infeld action \cite{Dirac:1962iy}, represents a specific instance of what is commonly referred to as nonlinear electrodynamics. It originated in the 1930s to resolve the issue of the electron's self-energy divergence in classical electrodynamics. This was achieved by imposing a limit on the electric field's magnitude at its source.}
Adding non-linear electrodynamics to $(2+1)$ Einstein gravity leads to the development of new solutions to the Einstein equations,
which can have important implications for our understanding of the universe.\\

Black hole physics, especially its thermodynamics, is among the most interesting and hot theoretical physics topics \cite{ter,ter1,ter2,ter3}. It is known that the thermodynamics of black
objects may be corrected due to thermal fluctuations \cite{1,Addazi:2021xuf}.
Such thermal fluctuations are significant when the black hole size reduces due to Hawking radiation. So, these
fluctuations may affect the thermodynamics of small black holes
effectively. As a result, it is interesting to study
thermodynamical stability of small black holes to find the final
stage of them. The lowest order of such corrections are
logarithmic function in the entropy formula \cite{2}, while higher
order corrections are also calculated \cite{2}.\newline
Following the path integral formulation in the Euclidean quantum gravity
quantization \cite{o, 01ab}, the temporal coordinate may be
rotated in a complex plane. Hence, the quantum gravitational
partition function, which is interpreted as statistical mechanics partition
function \cite{hawk} is given by%
\begin{eqnarray}
Z&=&\int [D]\exp (-I_{E})\nonumber\\
&=&\int_{0}^{\infty }dE\rho (E)\exp (-\beta E),
\label{Z}
\end{eqnarray}%
where $I_{E}$ is the Euclidean action, $\beta$ denotes the thermodynamic beta \cite{hawk,hawkis,Sakalli:2022xrb}, also known as coldness, is the reciprocal of the thermodynamic temperature of a system: $\beta ^{-1}\approx T$. In this regard, the equilibrium temperature is given by $T_{0}\approx \beta _{0}^{-1}$. In
addition, $\rho (E)$ stands for the density of states corresponding to above
statistical mechanics partition function, which is given by
\begin{equation*}
\rho (E)=\frac{1}{2\pi i}\int_{\beta _{0}-i\infty }^{\beta _{0}+i\infty
}d\beta \exp [S(\beta )],
\end{equation*}%
where the entropy $S$ is written in terms of both partition function and total
energy $E$, as $S=\beta E+\ln Z$. Neglecting the thermal fluctuations, one
can obtain $S_{0}=S(\beta )|_{\beta =\beta _{0}}$. Moreover, the quantum
fluctuations of a black hole increase when the size of the black hole
decreases, which yields an increasing of the thermal fluctuations in the
statistical mechanics partition function. As a result, there is no thermodynamic equilibrium at the Planck scale, whereas statistical fluctuations can be considered as thermal fluctuations in the partition function of Euclidean quantum gravity near the equilibrium in an intermediate regime, in which the black hole is large in comparison to the Planck scale but not large enough to ignore the quantum fluctuations. Therefore, using
the series expansion, we can find the corrected entropy $S(\beta )$ at a
temperature $\beta ^{-1}$ as a perturbation around the equilibrium
temperature $\beta _{0}^{-1}$ as follow%
\begin{equation}
S=S_{0}+\frac{1}{2}(\beta -\beta _{0})^{2}\left( \frac{\partial ^{2}S(\beta )%
}{\partial \beta ^{2}}\right) _{\beta =\beta _{0}}+\cdots ,  \label{a1}
\end{equation}%
where dots denote higher order corrections \cite{2}. It is worth noting that the
first derivative of entropy with respect to $\beta $ does not appear in Eq. (%
\ref{a1}), since we have $\left( \frac{\partial S(\beta )}{\partial \beta }%
\right) _{\beta =\beta _{0}}$ in equilibrium. Regarding Eq. (\ref{a1}), one
can rewrite the density of states as%
\begin{equation}
\rho (E)=\frac{\exp (S_{0})}{\sqrt{2\pi }}\left[ \left( \frac{\partial
^{2}S(\beta )}{\partial \beta ^{2}}\right) _{\beta =\beta _{0}}\right] ^{-%
\frac{1}{2}}.
\end{equation}%
and therefore, one can write
\begin{equation}
S=S_{0}-\frac{1}{2}\ln \left[ \left( \frac{\partial ^{2}S(\beta )}{\partial
\beta ^{2}}\right) _{\beta =\beta _{0}}\right].
\end{equation}
The mentioned calculations confirm that the thermal
fluctuations to the entropy of a black hole appear as $\ln
S_{0}T^{2}$. It is also known that such a logarithmic correction
to the entropy of a black object is obtained via different quantum gravity approaches, hence it seems to be a universal
 result, which does not depend on the given model. Indeed, such a logarithmic correction
has already been obtained using non-perturbative quantum general
relativity. Thus, the logarithmic corrected entropy can be prescribed by
\cite{4}
\begin{equation}
S=S_{0}-\frac{\alpha }{2}\ln {(S_{0}T^{2})},  \label{Corrected-Entropy}
\end{equation}%
where $S_{0}$ is the original entropy of the black hole without
thermal fluctuations, and the correction parameter $\alpha $ added
by hand to see the effect of logarithmic correction in the
analytical expressions.\newline When considering an ordinary thermodynamic system at equilibrium, statistical mechanics must be used to examine thermal fluctuations around the equilibrium. Assuming the black hole is  
the typical thermodynamic system which is in equilibrium, the same
statistical thermal fluctuations should be studied. It is worth
mentioning that the thermal fluctuation is an increasing function
of temperature, and thus, it becomes dominant at high temperatures.
Moreover, the temperature of the black hole increases when the
size of the black hole decreases due to the Hawking radiation \cite{o}.
Therefore, as a typical black hole becomes smaller, the effect of
quantum fluctuations would increase, and we could not neglect
their effects. The mentioned thermal fluctuations affect
thermodynamic variables, such as entropy and free energy, to
undergo possible correction. Following the classical argument
presented in Ref. \cite{1}, one can find that the entropy
correction arising from thermal fluctuations has a logarithmic
functional form. Here, we only considered the mentioned
logarithmic correction of entropy due to thermal fluctuations, and
in this study, we ignore the possible effect of one-loop
correction of massive graviton and matter fields.
\newline
Leading the thermal fluctuations to the semiclassical black hole
entropy primarily calculated as simple $\ln S_{0}$ by Kaul and
Majumdar \cite{5}. In that case, it has been suggested
that the coefficient of the correction may be universal \cite{7}.
Such a logarithmic correction to the rotating extremal black hole
entropy in four and five dimensions \cite{8} as well as
Schwarzschild and other non-extremal/non-rotating black holes in
different dimensions \cite{9} have been studied by Ashoke Sen. The
fact mentioned in \cite{9} is that at infinitesimal scales
(smaller than Planck scale), the manifold description of space-time
and hence, the equilibrium description of thermodynamics breaks
down. Therefore, the equilibrium corrected thermodynamics is
valid only at an intermediate scale. Therefore, we will consider
the system at such an intermediate scale, where statistical
fluctuations can be expressed as perturbations around equilibrium
thermodynamics. We indeed assume thermal fluctuations as small
perturbations around the equilibrium state. Hence, we can
use thermodynamics relations like that in equilibrium
thermodynamics. We should note that the one-loop effect of the
massive graviton and other matter fields \cite{9} is ignored in this
paper. The mentioned logarithmic correction has already been used to study simple regular black
hole solutions which satisfy the weak energy condition \cite{10}. It is recently
calculated from 3D black holes with soft hairy boundary
conditions \cite{new2}. It is worth mentioning that similar logarithmic
corrections to the black hole entropy were obtained from Kerr/CFT correspondence \cite{11}.

Recently, Eq. (\ref{Corrected-Entropy}) has been used to study
thermal fluctuations in a charged AdS black hole \cite{12}. In
addition, by using the Cardy formula, it has been found that
microscopic degrees of freedom of black objects governing by a
conformal field theory can also produce logarithmic corrections to the
entropy. So, it agrees with the universal behaviour of this logarithmic term. It may also be noted that even though the form of the
logarithmic correction seems to be a model-independent result; its
coefficient depends on the model parameters, and therefore, the
correction coefficient is model-dependent. In Ref. \cite{13}, a
black Saturn is considered and showed that the logarithmic
corrected entropy may
be given by%
\begin{equation}
S=S_{0}-\frac{\alpha }{2}\ln {(C_{0}T^{2})},  \label{Corrected-Entropy-22}
\end{equation}%
where $C_{0}$ is the original specific heat of the black Saturn. Then, it is
shown that both corrected entropies is given by (\ref%
{Corrected-Entropy}) and (\ref{Corrected-Entropy-22}) yields to the similar
results for the case of a black Saturn. The effect of the logarithmic correction on the
entropy functional formalism was also investigated in Ref. \cite{15}, and it is
found that the leading correction to the entropy in a microcanonical ensemble may be used
to recover modified theories of gravity like $F(R)$ theory \cite{17}. Moreover, the effects of thermal fluctuations on the thermodynamics of the modified
Hayward black hole are given in Ref. \cite{18}, and it is demonstrated that
the logarithmic correction helps to gain stability of the black hole. The
thermodynamics of a small singly spinning Kerr-AdS black hole studied in
Ref. \cite{19} and it is proven that the logarithmic correction of the form
given by Eq. (\ref{Corrected-Entropy}) becomes important when the size of a
black hole is sufficiently small. Since the black hole is a gravitating
system, its quantum correction investigation may give us a proposal to study
the quantum gravity effect with the possibility of its examination in the
lab. The effects of thermal fluctuations which are given in Eq. (\ref{Corrected-Entropy}) were studied for the 
Reissner-Nordstr\"{o}m-AdS black hole \cite{21} and shown that critical exponents are
the same as critical exponents without thermal fluctuations. It is also
possible to study critical phenomena; as an example, $P-V$ criticality of
the leading-order entropy-corrected AdS black holes in massive gravity theory has
been recently studied \cite{22}.

On the other hand, the logarithmic corrected entropy may improve
some hydrodynamic quantities as well as thermodynamic ones and
such an improvement may be important from the AdS/CFT point of
view \cite{23}. In Ref. \cite{23}, it has been demonstrated that
the lower bound of shear viscosity to entropy ratio
\cite{24,27,29} due to logarithmic correction may be violated.
The effects of thermal fluctuations (see Eq.
(\ref{Corrected-Entropy})) on a regular black hole of non-minimal
Einstein-Yang-Mill theory with the gauge field of magnetic Wu-Yang
type and a cosmological constant were studied in \cite{new}.
So, considering different black hole solutions with such
thermal fluctuations can help us to take a deeper look at the
applications of AdS/CFT correspondence.

In addition to the motivations stated above, the logarithmic
corrected entropy may be interested in various cosmology topics,
like the logarithmic entropy-corrected holographic dark energy
models \cite{30}. As is well known, logarithmic corrections arise due to quantum effects, and they have important implications for how
the gravitational field interacts with other fields in the universe. Besides, the
logarithmic corrected thermodynamics of charged BTZ black hole obtained in the
Einstein-Maxwell gravity was studied in \cite{32,Nashed:2020kdb}. Now, we would like to study
the corrected thermodynamics (\ref{Corrected-Entropy}) for the
Einstein-Born-Infeld solution of the charged BTZ black hole in massive
gravity \cite{main}. 
The graviton
mass has bounded to ${m_g} \le 7.7 \times {10^{ - 23}}eV/{c^2}$ using the observation of the gravitational wave.
\cite{LIGO2017}. Besides, the mass of graviton is typically bounded to be a few
times the current value of the Hubble parameter, i.e., ${m_g} \le {10^{ -
30}-10^{ - 33}}eV/{c^2}$, in which for graviton mass region $
m_{g}\ll 10^{ - 33}eV/{c^2}$ , its observable effects would be
undetectable \cite{dRGT,deRhamREVIEW2014,Cai:2013lqa,Cai:2014upa,Gannouji:2013rwa}.
It was also discovered that nonlinear charged BTZ-mass black holes produce black remnants \cite{main}. {At this point it is worth noting that the Born-Infeld source introduces nonlinearity
to the equations, and in certain conditions, this nonlinearity can influence the asymptotic behavior
of the solution \cite{Yan:2018yyz}. One possible modification involves cases where the Born-Infeld contribution becomes
dominant in regions with extreme curvature, leading to deviations from the asymptotic behavior predicted
by standard models \cite{Garrione}. Such modifications could have implications for the understanding of gravitational
interactions in regimes characterized by strong fields. However, it is crucial to note that our decision
to emphasize cases with unaltered asymptotic behavior does not negate the relevance of exploring
modifications. The chosen focus allows us to delve deeply into specific aspects of the problem and
draw attention to the nuanced effects of the Born-Infeld source. Future investigations may naturally
extend to scenarios where asymptotic modifications become a central theme, and we encourage further
exploration in this direction. In summary, while our current analysis concentrates on cases with unaltered
asymptotic behavior to underscore specific features, we recognize the significance of considering and
discussing scenarios where modifications may occur.} 

In this paper, we mainly investigate the effects of thermal fluctuations on the (non)linearly charged BTZ
black hole and study the behavior of this black hole at a small horizon
radius. Thus, we aim to obtain the quantum effects. At that scale, the black hole
entropy should be corrected by logarithmic term (in leading order)
coming from statistical fluctuations, and hence, the black hole
thermodynamics is modified by reducing black hole size. This situation will also affect the
black hole stability in small areas. In the next section, Sec. \ref{sec2}, we first review the Einstein-Maxwell
solutions under the effects of logarithmic corrections. In Sec. \ref{sec3},
we investigate the effects of logarithmic corrections on the
thermodynamics of the Einstein-Born-Infeld solutions as
nonlinearly charged black hole solutions. We shall also discuss the
critical points and compare the results of both solutions. Section
\ref{sec4} is devoted to the study of geometrical thermodynamics. Finally,
in Sec. \ref{sec5}, we draw our conclusions.

\section{Thermodynamics of the Einstein-Maxwell solution} \label{sec2}

In this section, we introduce the model of our interest in this
paper. As is well-known, $3$-dimensional Einstein-massive gravity in
the presence of abelian electrodynamics, as a matter field, is described by the following action
\cite{dRGT,deRhamREVIEW2014,Cai:2012db,Vegh2013}
\begin{equation}
I=-\frac{1}{16\pi }\int d^{3}x\sqrt{-g}\left[ \mathcal{R}-2\Lambda +L(%
\mathcal{F})+m^{2}\sum_{i=1}c_{i}\mathcal{U}_{i}(g,f)\right] ,
\label{action}
\end{equation}%
where $\mathcal{R}$ is the scalar curvature, $\Lambda $ is the cosmological
constant, $L(\mathcal{F})$ is an arbitrary Lagrangian of electrodynamics, and
the last term contains the potential terms of massive gravity. Indeed, $%
m$ denotes the massive parameter, $c_{i}$'s are some constants and
interaction terms $\mathcal{U}_{i}$'s are symmetric polynomials of the
eigenvalues of $d\times d$ matrix $\mathcal{K}_{\,\,\,\,\,\nu }^{\mu }=\sqrt{%
{g^{\mu \alpha }}{f_{\alpha \nu }}}$ in an arbitrary $d$-dimensions. It is worth pointing out that in a $d$-dimensional spacetime, the interaction terms $%
\mathcal{U}_{i}$ have no contribution for $i>d-2$, and therefore,
there is only one potential term of massive gravity in three
dimensional spacetime as
${\mathcal{U}_{1}}=\mathcal{K}_{\,\,\,\,\mu }^{\mu }$ (see
\cite{dRGT,deRhamREVIEW2014,Cai:2013lqa,Cai:2014upa,Cai:2012db,Vegh2013,Cai2015} for more details).
Considering the action (\ref{action}) and applying the variational
principle, the field equations can be written as
\begin{equation}
{R_{\mu \nu }}-\frac{{g_{\mu \nu }}}{2}\left( R-2\Lambda \right) -\frac{{%
m^{2}{c_{1}}}}{2}({\mathcal{U}_{1}}{g_{\mu \nu }}-{\mathcal{K}_{\mu \nu }})=%
\frac{1}{2}g_{\mu \nu }L(\mathcal{F})-2\frac{dL(\mathcal{F})}{d\mathcal{F}}%
F_{\mu \lambda }{F_{\nu }}^{\,\lambda },  \label{FE1}
\end{equation}%
\begin{equation}
{\nabla _{\mu }}\left( \frac{dL(\mathcal{F})}{d\mathcal{F}}{F^{\mu \nu }}%
\right) =0,  \label{FE2}
\end{equation}%
Here, we intend to obtain static-charged black hole solutions of
massive gravity in three dimensions. To do so, we should consider
an appropriate ansatz for the auxiliary reference metric $f_{\mu
\nu }$ (nonphysical metric) and a $3$-dimensional line-element
ansatz to describe the physical metric $g_{\mu \nu }$
\cite{deRhamREVIEW2014,Vegh2013,Cai2015}:
\begin{equation}
{f_{\mu \nu }}=diag\left( {0,0,c_{0}^{2}}\right) ,  \label{reference metric}
\end{equation}%
\begin{equation}
g{_{\mu \nu }}=diag\left( -f(r){,}\frac{{1}}{f(r)}{,r^{2}}\right) ,
\label{metric1}
\end{equation}%
where $c_{0}$ is a positive constant. Now, we would like to review the
Einstein-Maxwell solution and then consider some corrected
thermodynamics quantities. Regarding the Lagrangian of Maxwell
field ($L(\mathcal{F})=-\mathcal{F}$) with the aforementioned field
equations, one can find the metric function as \cite{main}
\begin{equation}
f(r)=-\Lambda r^{2}-m_{0}-2q^{2}\ln {(\frac{r}{l})}+m^{2}cc_{1}r.
\label{f(r)}
\end{equation}%
In the above equation, two integration constants $m_{0}$ and $q$ are related to
the mass $M$ and electric charge $Q$\ of black hole as $m_{0}=8M$ and $q=2Q$
(see \cite{main} for more details). In addition, the massive parameter
is denoted by $m$, while $c$ and $c_{1}$ are two positive constants. It is
clear that the linearly charged BTZ black hole solution is obtained by setting $%
m=0$. Considering the behaviour of the metric function, one finds that
it is possible to have two horizons $r_{+}$ (event horizon) and $r_{-}$ (inner horizon). Since we
would like to study the black hole thermodynamics at the event horizon, we
calculate all quantities at $r_{+}$, which is the largest real positive root
of the metric function \eqref{f(r)}.\newline
Evaluating the action of this configuration diverges at
spatial infinity. To overcome this problem, one must
consider supplementary boundary terms in addition to the bulk action. These
$2$-dimensional boundary actions (called Gibbons-Hawking and counter-term
actions) do not change the field equations arising from the variational
principle. As a result, considering the obtained black hole solutions with
AdS asymptote and inspired by AdS/CFT correspondence, one would use the
counter-term method to calculate the finite mass of a black hole (see \cite{A}
for more details). So, after some straightforward calculations, we obtain $M=\frac{m_{0}}{8}$%
. In addition, one can use the standard ADM mass calculation and rewrite the
Einstein's action in ADM form to obtain the finite mass, as discussed in Ref.
\cite{B}. At this stage, it is worth noting that in Ref. \cite{B} the natural units of
$8G=1$ were used; however in this work, we follow $G=1$. It is worth mentioning
that the result of finite mass in Ref. \cite{A} is valid here since the
dominant term of the metric function is $\Lambda r^{2}$-term at spatial
infinity ($r\rightarrow \infty$).\newline
The following dimensionless mass is obtained from the fact that the metric
function is zero at the event horizon%
\begin{equation}
M=-\frac{\Lambda }{8}r_{+}^{2}-Q^{2}\ln {(\frac{r_{+}}{l})}+\frac{m^{2}cc_{1}%
}{8}r_{+}.  \label{Mass}
\end{equation}%
Applying the surface gravity interpretation, we can obtain the following
relation for the temperature%
\begin{equation}
T=\frac{-\Upsilon }{4\pi r_{+}},  \label{T}
\end{equation}%
where $\Upsilon\equiv2\Lambda r_{+}^{2}-m^{2}cc_{1}r_{+}+8Q^{2}$. Also, the uncorrected black hole entropy is given by the area law \cite{Hawking6}%
\begin{equation}
S_{0}=\frac{\pi }{2}r_{+}.  \label{Ent1}
\end{equation}%
To avoid the dimensional problem of the logarithm argument in three
dimensions, we use the following dimensionless thermodynamic quantities of the
charged BTZ black hole, which are re-defined based on different powers of the
length scale 
\begin{equation}
T^{\prime }=lT,\text{ \ }S_{0}^{\prime }=S_{0}/l.  \label{dimensionless}
\end{equation}
To have a positive temperature, one has to regard $\Upsilon <0$ as an
essential condition.\\
Therefore, the logarithmic corrected (dimensionless) entropy (\ref
{Corrected-Entropy}) can be obtained as 
\begin{equation}
S^{\prime }=S_{0}^{\prime }-\frac{\alpha }{2}\ln {(S}^{\prime }{_{0}T}%
^{\prime }{^{2})}=\frac{\pi r_{+}}{2l}-\frac{\alpha }{2}\ln {\left( \frac{%
l\Upsilon ^{2}}{32\pi r_{+}}\right) }.  \label{Corrected-Entropy1}
\end{equation}%
It should be noted that the dimensionless parameter $\alpha $ is added by hand
to trace the effect of logarithmic correction in analytical expressions.
Hence, its value should be $\alpha=0$ (uncorrected) or $\alpha=1$
(corrected). However, we can consider it as a free parameter and investigate
other possibilities. The first law of black hole thermodynamics is written
as%
\begin{equation}
dM=T^{\prime }dS^{\prime }+\Phi dQ,  \label{FirstL1}
\end{equation}%
where $\Phi $ denotes the electric potential, which is calculated using the
the electric potential at the event horizon \cite{main}%
\begin{equation}
\Phi =-2Q\ln {(\frac{r_{+}}{l}).}
\end{equation}%
Hence, the Smarr formula is given by
\begin{equation}
M=2T^{\prime }S^{\prime }+\Phi Q.  \label{Smarr1}
\end{equation}%
It has been found that the first law of thermodynamics (\ref{FirstL1}) is
valid only for the case of $\alpha =0$ (uncorrected entropy). However, in
the presence of the logarithmic correction, we can examine the validity of
the first law of black hole thermodynamics by choosing the appropriate
parameters of the solutions. In that case (corrected entropy), the
(dimensionless) internal energy is given by
\begin{eqnarray}
E &=&\int {T}^{\prime }{dS}^{\prime }  \notag  \label{Int-E1-1} \\
&=&\left[ \frac{m^{2}cc_{1}}{8}-\frac{\Lambda r_{+}}{8}\right]
r_{+}-Q^{2}\ln {(\frac{r_{+}}{l})}+\frac{\alpha l}{8\pi }\left[ 6\Lambda
r_{+}+\frac{8Q^{2}}{r_{+}}-m^{2}cc_{1}\ln {(\frac{r_{+}}{l})}\right] .
\end{eqnarray}%
It is obvious that $\alpha$ decreases the value of $E$. In addition, one can obtain the dimensionless heat
capacity as follow%
\begin{equation}
C=T^{\prime }\frac{dS^{\prime }}{dT^{\prime }}=\frac{\left( \frac{\pi r_{+}}{%
l}-\alpha \right) \Upsilon }{4l\left( \Lambda r_{+}^{2}-4Q^{2}\right) }-%
\frac{\alpha }{l}.  \label{S-Heat1-1}
\end{equation}%
Regarding the mentioned condition ($\Upsilon <0$), in the case of $\alpha =0$, we have
thermodynamical stability when $\Lambda r_{+}^{2}-4Q^{2}<0$. The asymptotically AdS black hole is thermally stable.

\begin{figure}[h!]
\begin{center}
$%
\begin{array}{cccc}
\includegraphics[width=65 mm]{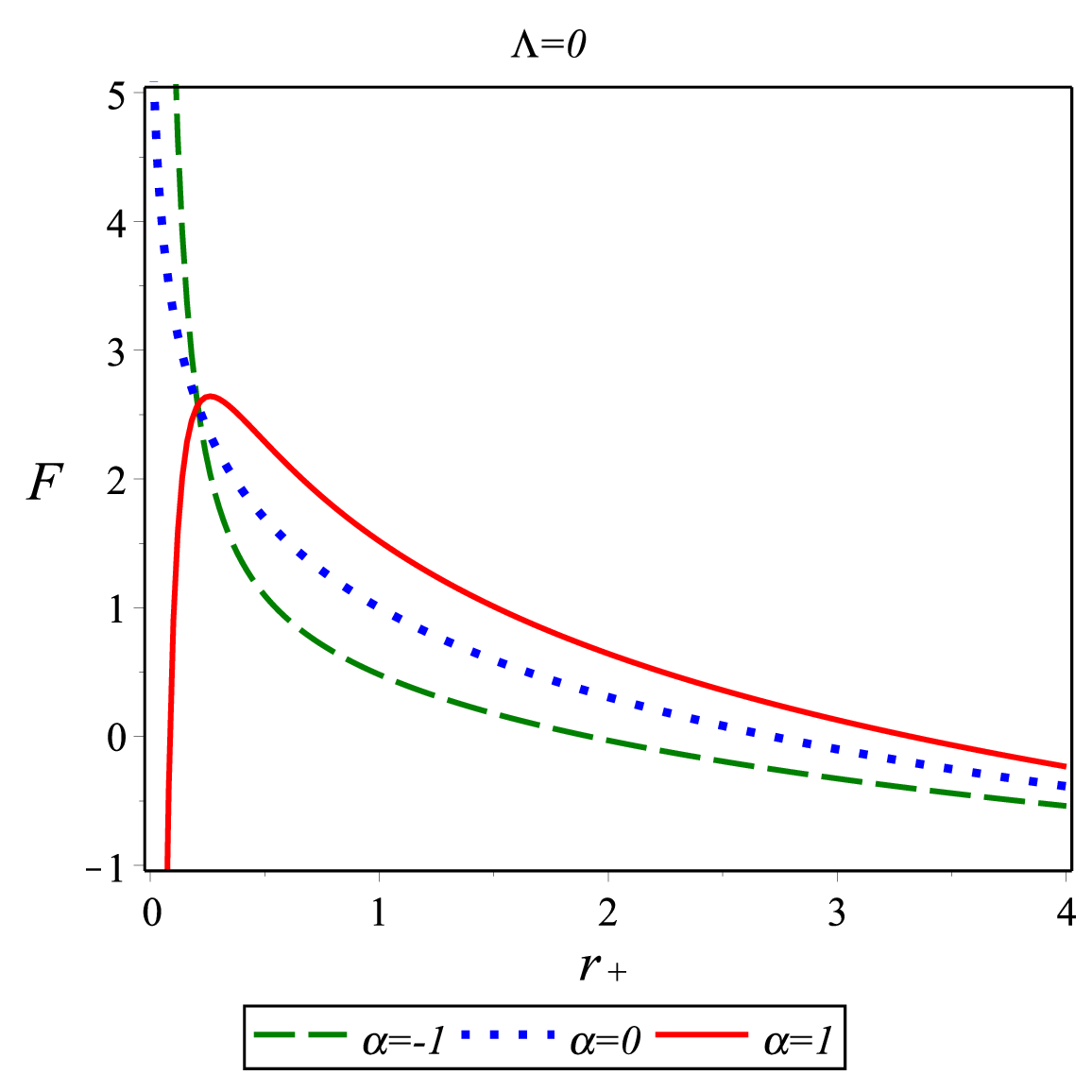}%
\includegraphics[width=65 mm]{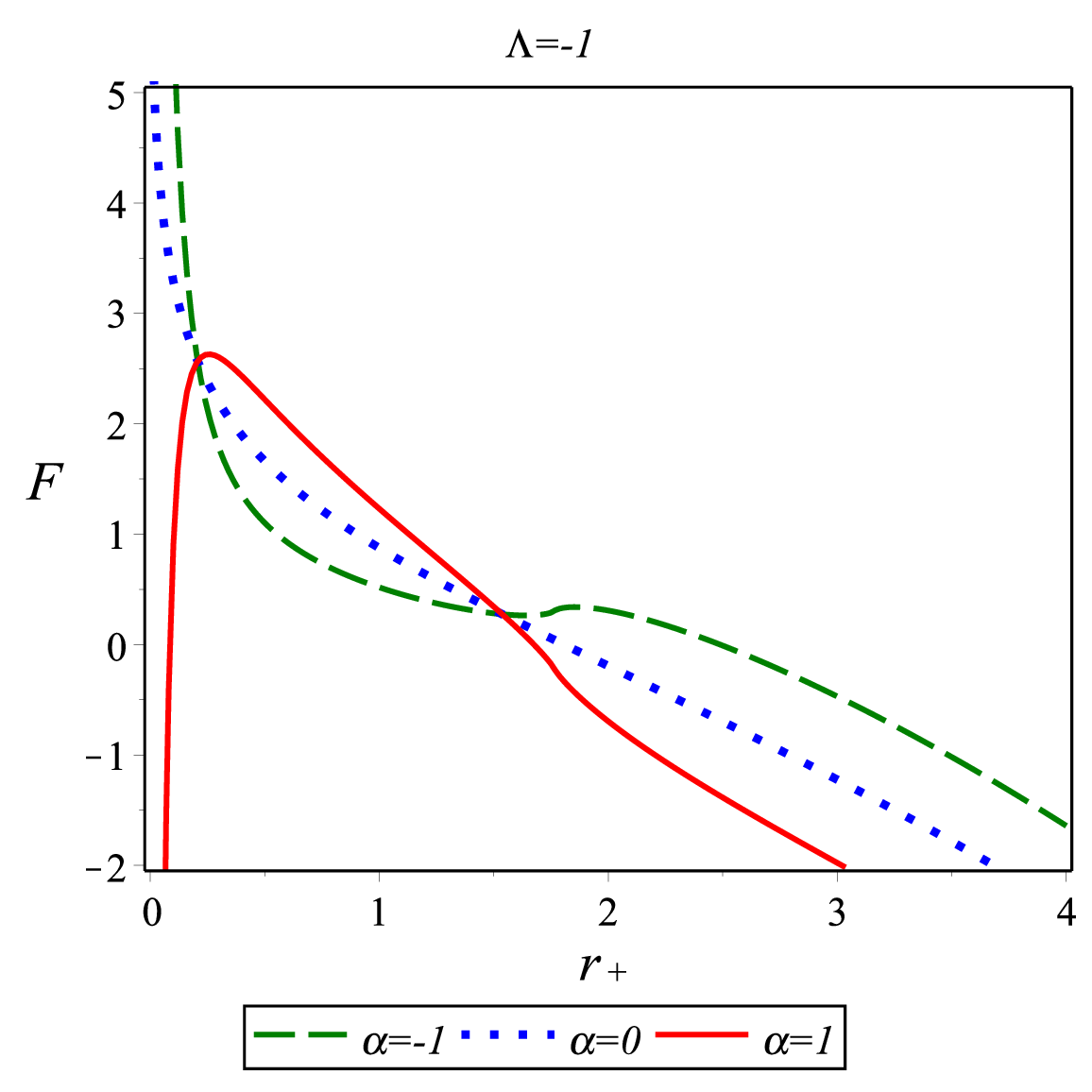} &  &  &
\end{array}%
$%
\end{center}
\caption{Helmholtz free energy in terms of $r_{+}$ for unit value of the
model parameters.}
\label{fig1}
\end{figure}

Now, we would like to discuss two cases of $\alpha >0$ and $\alpha <0$ by
respecting the condition of having a positive temperature. It should be
noted that to recover Einstein's gravity when we turn off the massive gravity
couplings, the $\alpha$ parameter should be positive. In the case of a
negative $\alpha$, the situation is the same as vanishing $\alpha$. In
In other words, the AdS black hole is thermally stable. However, for
positive $\alpha$, the situation is not trivial. In this special case ($%
\alpha >0$), we should consider the following restriction to obtain stable
solutions%
\begin{equation}
\begin{array}{cc}
\alpha >\frac{\pi r_{+}}{l}\left( 1-\frac{4\left( \Lambda
r_{+}^{2}-4Q^{2}\right) }{6\Lambda r_{+}^{2}-m^{2}cc_{1}r_{+}-8Q^{2}}\right)
, & \text{for }r_{+}<\frac{2Q}{\sqrt{|\Lambda| }}%
 , \\
&  \\
\alpha <\frac{\pi r_{+}}{l}\left( 1-\frac{4\left( \Lambda
r_{+}^{2}-4Q^{2}\right) }{6\Lambda r_{+}^{2}-m^{2}cc_{1}r_{+}-8Q^{2}}\right)
, & \text{for }r_{+}>\frac{2Q}{\sqrt{|\Lambda| }}.%
\end{array}
\label{Condition}
\end{equation}%
\newline
As we mentioned already, the first law of thermodynamics is only valid under
the special condition ($\alpha =0$). Near the equilibrium, we use the
following expression%
\begin{equation}
6\Lambda r_{+}^{2}-m^{2}cc_{1}r_{+}-8Q^{2}=\epsilon ,  \label{condition1-1}
\end{equation}%
where $\epsilon $ is an infinitesimal parameter. In the case of $\epsilon =0$%
, we have a complete equilibrium with the validity of the second law. In
that case, the specific heat is reduced to the following expression 
\begin{equation}
C=\frac{\left( \frac{\pi r_{+}}{L}-\alpha \right) \epsilon }{4\left( \Lambda
r_{+}^{2}-4Q^{2}\right) }-\frac{\pi r_{+}}{L}.
\end{equation}
For the cases of $\alpha=0$ and $\alpha=\pm1$, we find specific heat is
negative, hence the black hole is unstable. There is a singular point as
\begin{equation}
r_{+c}=\frac{2Q}{\sqrt{|\Lambda| }}.  \label{r+c1}
\end{equation}
which is not a phase transition point because the black hole is unstable
before and after that point.\newline
We can investigate the Helmholtz free energy via the following relation 
\begin{equation}
F=E-TS.  \label{Helmhultz1}
\end{equation}%
Then, using Eqs. (\ref{T}), (\ref{Corrected-Entropy1}) and (\ref{Int-E1-1}),
one can obtain the Helmholtz free energy as follows
\begin{eqnarray}
F &=&-\frac{\alpha l}{8\pi r_{+}}\left[ \Upsilon \ln \left( \frac{l\Upsilon
^{2}}{32\pi r_{+}}\right) -6\,\Lambda r_{+}^{2}+cc_{1}{m}^{2}r_{+}\ln \left(
\frac{r_{+}}{l}\right) -8{Q}^{2}\right]  \notag \\
&&+\frac{\Lambda r_{+}^{2}}{8}+{Q}^{2}\left[ 1-\ln \left( \frac{r_{+}}{l}%
\right) \right] \text{ \ }  \label{Helmholtz}
\end{eqnarray}%
In the plots of Fig. \ref{fig1} we can see the typical behaviour of the Helmholtz
free energy in terms of $r_{+}$.\newline
The case of $\alpha=0$ (absence of thermal fluctuations) is approximately
similar for all cases of $\Lambda=0, -1$. We can see that there is a
special radius where the effect of thermal fluctuations vanishes. In the
case of $\Lambda=-1$, we have two separate radii where the effect of thermal
fluctuations vanishes. In the presence of thermal fluctuations, the
Helmholtz free energy has a maximum, which
may indicate an instability or phase transition. It needs further study, which will be done via specific heat.

\section{Thermodynamics of the Einstein-Born-Infeld solution} \label{sec3}

\begin{figure}[h!]
\begin{center}
$%
\begin{array}{cccc}
\includegraphics[width=80 mm]{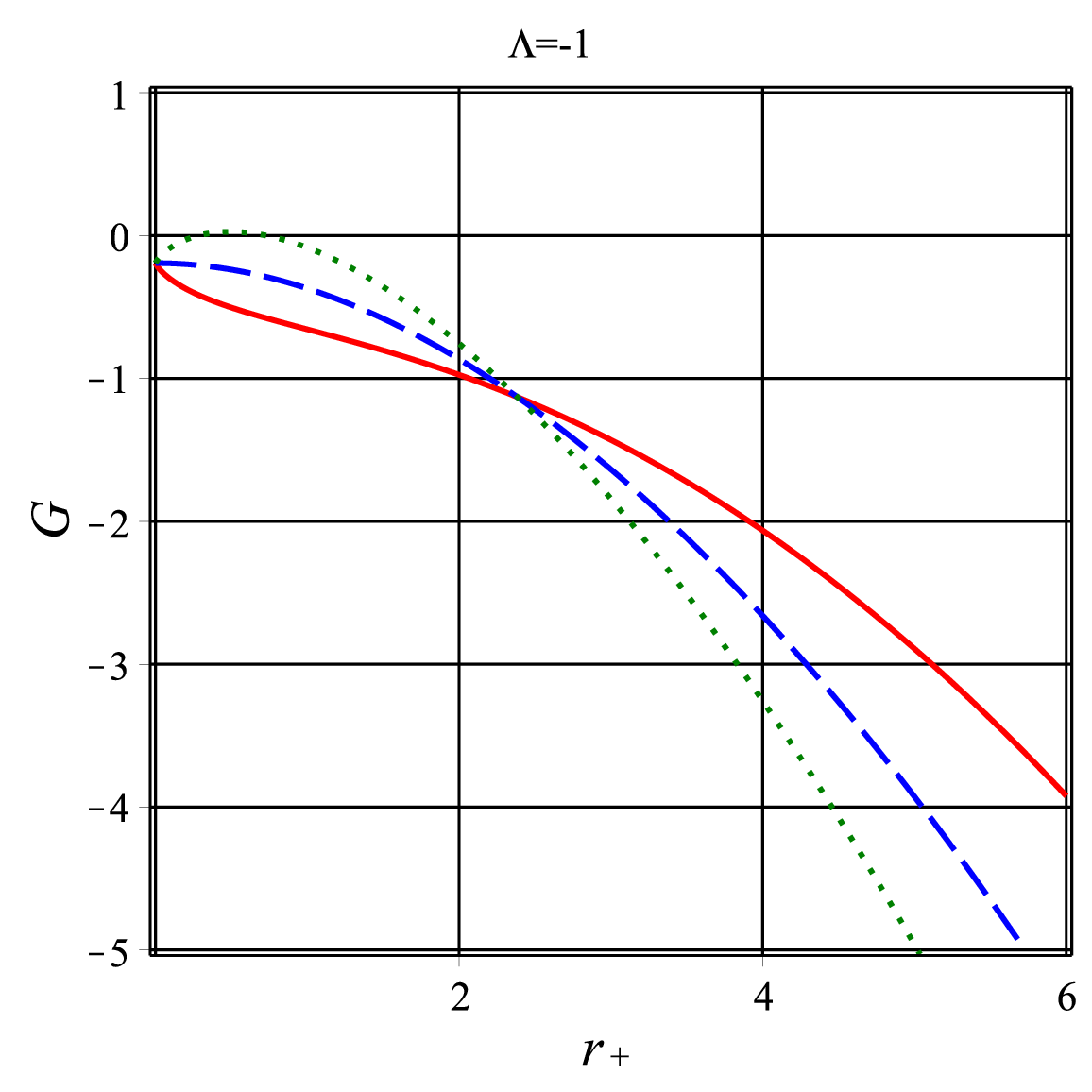}
&  &  &
\end{array}%
$%
\end{center}
\caption{Gibbs free energy in terms of $r_{+}$ for $m=2$ and $\protect\beta%
=0.5$; we set unit values for all other parameters. Blue dashed lines
represented the case of $\protect\alpha=0$, red solid lines represented the
case of $\protect\alpha=1$ and green dotted lines represented the case of $%
\protect\alpha=-1$.}
\label{fig3}
\end{figure}

Now, we will generalize our discussion to the case of
nonlinear electrodynamics. So, we regard the same field equations
as presented in Eqs. (\ref{FE1}) and (\ref{FE2}) with the
following Lagrangian of Born-Infeld theory
\begin{equation*}
L(\mathcal{F})=4\beta ^{2}\left( 1-\sqrt{1+\frac{\mathcal{F}}{2\beta ^{2}}}%
\right),
\end{equation*}
where $\beta $ is the nonlinearity parameter (notably, for $\beta
\longrightarrow \infty $ the Maxwell case is recovered). In this
case, the nonlinearly charged AdS black hole is given by the metric
(\ref{metric1}) with the following metric function
\begin{equation}
f(r)=-\Lambda r^{2}-m_{0}+2\beta ^{2}r^{2}(1-\Gamma )+q^{2}\left[ 1-2\ln
\left( {\frac{r}{2l}(1+\Gamma )}\right) \right] +m^{2}cc_{1}r,  \label{f2}
\end{equation}%
where $\beta $ is nonlinearity parameter and%
\begin{equation}
\Gamma =\sqrt{1+\frac{q^{2}}{r^{2}\beta ^{2}}}.
\end{equation}%
Since the asymptotic behavior of this nonlinearly charged solution is the
same as that of linearly charged one, we find that $m_{0}=8M$ and $q=2Q$,
as before. Since for very large $r$, the metric function of
BI-BTZ-AdS black hole reduces to Maxwell-BTZ-AdS solution; one can find that
the BI contribution does not change the asymptotic behavior of the
solutions. Therefore, the previous finite mass discussion is also valid here
and we can write $m_{0}=8M$. In the case of $f(r=r_{+})=0$, one can obtain%
\begin{equation}
M=-\frac{\Lambda }{8}r_{+}^{2}-Q^{2}\ln \left( {\frac{r_{+}}{l}(1+\Gamma
_{+})}\right) +\frac{2r_{+}^{2}\beta ^{2}(1-\Gamma
_{+})+4Q^{2}+m^{2}cc_{1}r_{+}}{8},
\end{equation}%
where $\Gamma _{+}=\Gamma |_{r=r_{+}}$. The original black hole entropy of
this case is similar to Eq. (\ref{Ent1}) while the Hawking temperature is
given by%
\begin{equation}
T=\left. \frac{f^{\prime }(r)}{4\pi }\right\vert _{r=r_{+}}=-\frac{\Lambda
r_{+}}{2\pi }-\frac{4Q^{2}}{\pi r_{+}(1+\Gamma _{+})}+\frac{m^{2}cc_{1}}{%
4\pi }.  \label{Temp2}
\end{equation}%
In addition, the electric potential of the solution is given by \cite{main}%
\begin{equation}
\Phi =-2Q\ln \left( {\frac{r_{+}}{2l}(1+\Gamma _{+})}\right) .  \label{U1}
\end{equation}%
It is obvious that the first law (\ref{FirstL1}) is valid for the case of $%
\alpha =0$. In the presence of $\alpha $, we also have the validity of the first law
of black hole thermodynamics under the following conditions (in units of $l$),
\begin{eqnarray}
&-&6\,{\beta }^{2}\Lambda \,\left( \Gamma _{+}^{3}+2\right) r_{+}^{4}+{\beta
}^{2}{m}^{2}cc_{1}\,\left( \Gamma _{+}^{3}+2\right) r_{+}^{3}+8\,{Q}^{2}{m}%
^{2}cc_{1}r_{+}-64\,{Q}^{4}  \notag  \label{condition2} \\
&+&\left( 16\,{\beta }^{2}\left( \frac{1}{16}{m}^{2}cc_{1}r_{+}-\frac{3}{8}%
\Lambda \,r_{+}^{2}+{Q}^{2}\right) \Gamma _{+}+16\,{Q}^{2}\left( {\beta }%
^{2}-3\,\Lambda \right) +1\right) r_{+}^{2}=0,
\end{eqnarray}%
which satisfied at $r_{+}=r_{+}^{\ast }$, meaning that, under effect of
logarithmic correction, the first law of black hole thermodynamics, is only valid at radius $r_{+}^{\ast}$, an analogous behavior to the linear case.\newline
Now, we can study Gibbs free energy graphically by Fig. \ref{fig3}. We can see the typical behavior of the
Gibbs free energy in terms of $r_{+}$ by variation of $\alpha $ to see the
effect of logarithmic correction. We can
find a special radius where the logarithmic correction does not affect the
Gibbs free energy. Hence, the effect of thermal fluctuations strongly
depends on the $r_{+}$.

\begin{figure}[h]
\begin{center}
$%
\begin{array}{cccc}
\includegraphics[width=62 mm]{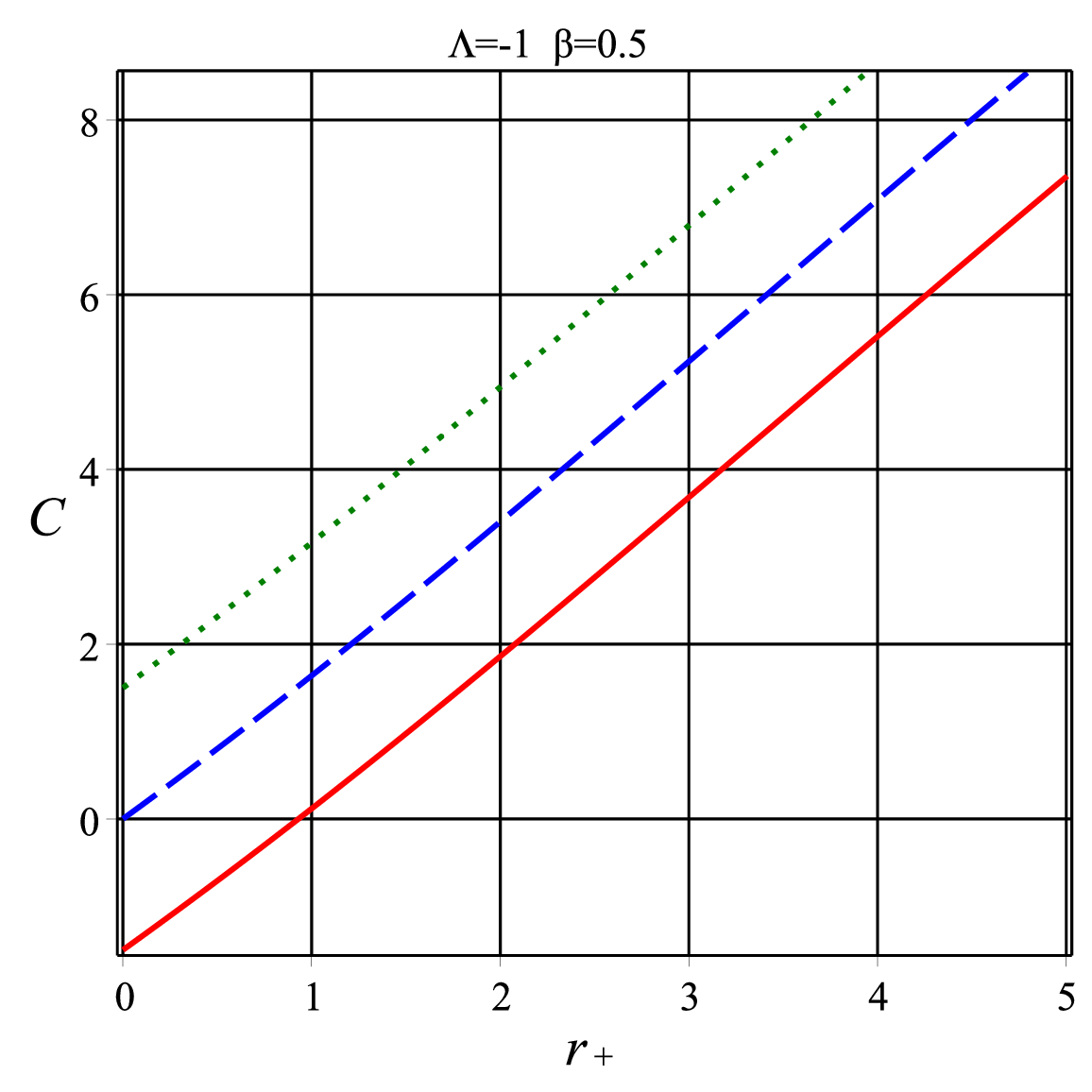}\includegraphics[width=62 mm]{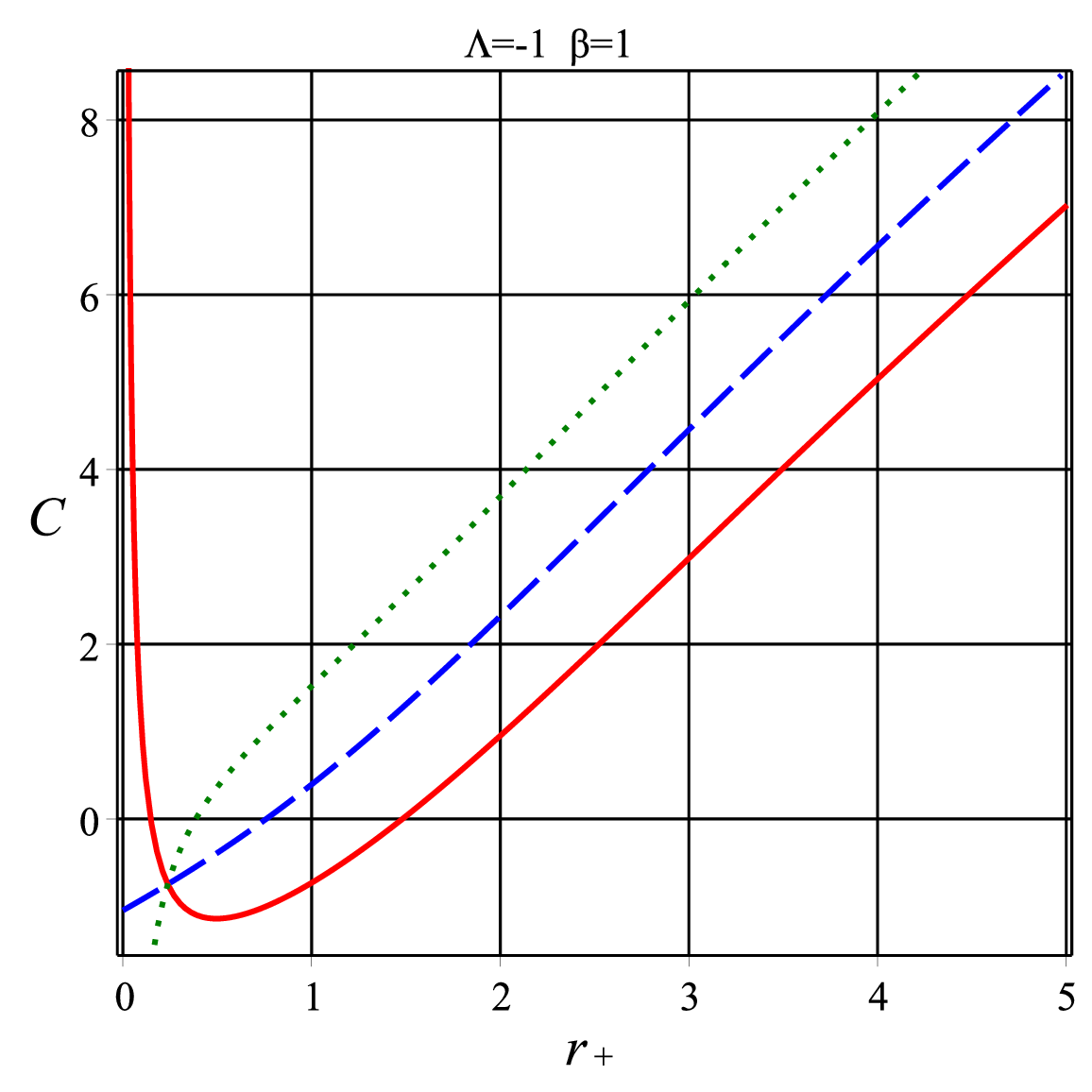}
&  &  & 
\end{array}%
$%
\end{center}
\caption{Specific heat in terms of $r_{+}$ for $m=2$ and we set unit values
for all other parameters. Blue dashed lines represented the case of $\protect%
\alpha =0$, red solid lines represented the case of $\protect\alpha =1$ and
green dotted lines represented the case of $\protect\alpha =-1$.}
\label{fig4}
\end{figure}

Now, we can study the stability of the black hole by using heat capacity given
by the general equation $C=T^{\prime }\frac{dS^{\prime }}{dT^{\prime }}$.
Results are illustrated by plots of Fig. \ref{fig4}.

According to the plots of Fig. \ref{fig4}, we can observe the behavior of
specific heat. In the case of $\beta =0.5$, we can see
linear behavior of specific heat with a
completely positive value for $\alpha =-1$ as in the previous section. The denominator of specific heat is not affected by the
logarithmic correction and gives the following critical radius%
\begin{equation}
r_{+c}=\pm \frac{2(\Lambda -2\beta ^{2})Q}{\beta \sqrt{\Lambda (4\beta
^{2}-\Lambda )}}.  \label{r+c2}
\end{equation}%
Discussion and details about this relation are completely similar to those
reported in Ref. \cite{main}. In the case of $\alpha =0$ (dashed lines), we can see the linear behavior
of specific heat as discussed in Ref. \cite{main}. However, for $\beta =0.5$, the logarithmic correction with $\alpha =1$ leads to a decreasing value of
specific heat, while $\alpha =-1$ increases its value, and hence, the black
hole is completely stable. In the case of $\beta =1$ as discussed by Ref.
\cite{main}, the specific heat may be negative at $\alpha =0$ corresponding
to small $r_{+}$. However, there is no asymptotic behavior, and hence, there
is no phase transition. In the presence of the thermal fluctuations with $\alpha
=-1$, we can see a large/small phase transition at $r_{+}\approx 0.4$ through a first-order phase transition
(dotted line right plot of Fig. \ref{fig4}). For $\alpha =1$, there
is a minimum of specific heat at the negative region, and therefore, the
specific heat has two zeros for this case.

In the next section, we investigate the critical behavior of both linear and nonlinear solutions.

\subsection{Critical points}
To discuss the critical point, we should study the thermodynamics pressure.

Thermodynamic pressure in the context of BTZ thermodynamics has been extensively studied in the literature. For example, non-rotating BTZ black holes are considered by Ref. \cite{P1} and found that there is no critical point and holographically dual Van der Waals behavior. The same result already obtained for BTZ black holes in nonlinear electrodynamics \cite{P2}. Now, we show the same result for the (non)linearly charged BTZ
black hole in massive gravity.

First of all, mention that the thermodynamics pressure $P$ is related to the cosmological constant via,
\begin{equation}
P=-\frac{\Lambda }{8\pi }. \label{pressure}
\end{equation}
Its conjugate quantity is the thermodynamics volume which is given by
\begin{equation}
V=\frac{4}{3}\pi r_{+}^{3}. \label{Vol}
\end{equation}
In that case, it is possible to investigate the critical points and phase
transition like a Van der Waals fluid.\newline
In the Einstein-Maxwell solution, combining condition (\ref{r+c1}) with the
pressure (\ref{pressure}) gives \cite{main}%
\begin{equation}
P_{M}=-\frac{Q^{2}}{2\pi r_{+}^{2}},  \label{pressure1}
\end{equation}%
then using condition (\ref{condition1-1}) with $\epsilon =0$ to have
validity of the first law of thermodynamics in the presence of logarithmic
correction, we obtain%
\begin{equation}
P_{M}={\frac{{m}^{4}c^{2}c_{1}^{2}}{2^{11}\,\pi \,{Q}^{2}}}.
\label{pressure1-1}
\end{equation}%
We can see that there is no extremum, therefore there is no Van der Waals-like behavior, as well as the case of $\alpha
=0$. It means that there is no holographic dual Van der Waals fluid. Similar results may obtained
for Einstein-Born-Infeld solution where \cite{main}%
\begin{equation}
P_{BI}=\pm \frac{\beta ^{2}Q^{2}}{\pi \left[ 4Q^{2}+\beta
^{2}r_{+}^{2}(1+\Gamma _{+})\right] }.  \label{pressure2}
\end{equation}%
Obtaining $r_{+}$ from (\ref{condition2}) and inserting it in the $P_{BI}$,
we find that there is no extremum. Hence, we can see there is no critical
points and Van der Waals-like behavior. However, if we neglect the logarithmic correction, which is a good approximation for the large black hole, we may find holographic dual Van der Waals fluid \cite{22}.

\section{Geometrical thermodynamics} \label{sec4}

\label{Geometrical thermodynamics}

Geometrical thermodynamics is one of the interesting approaches to
investigating the thermodynamic structure of black holes. In this approach,
one may build a thermodynamic phase space by considering mass or entropy as
a potential, which is a function of its corresponding extensive parameters.
Regarding the thermodynamic metric in such a phase space, the corresponding
Ricci scalar carries information about thermodynamical properties. In
other words, it is shown that the divergence points of thermodynamical
Ricci scalar coincides with the bound point or associated phase transition
points. There are different geometrical methods so far; Weinhold \cite%
{WeinholdI,WeinholdII}, Ruppeiner \cite{RuppeinerI,RuppeinerII},
Quevedo \cite{QuevedoI,QuevedoII} and HPEM \cite{HPEMI,HPEMII}.
Such methods and a comparative study are considered in different
researches in the
context of black holes and superconductors \cite%
{HanC,BravettiMMA,Ma,GarciaMC,ZhangCY,MoLW2016,BasakCNS,comp}.
Taking into account the thermodynamical concepts, it is obvious
that no ensemble dependency should exist. The existence of
ensemble dependency or anomalies in the geometrical thermodynamic
models \cite{dependency} may be observed as: I) characteristic
points (bound point and/or phase transition points) are not
matched with divergence point(s) of thermodynamical Ricci scalar.
II) extra divergence point is observed for the Ricci scalar which
is not matched with any characteristic points. Although, Quevedo
introduced an interesting Legendre invariant method to remove the
problems of Weinhold and Ruppeiner approaches, it is shown that
the Quevedo metric may encounter an anomaly
\cite{HPEMI,HPEMII}. The Legendre invariant HPEM metric which is
conformally related to the Quevedo metric was proposed to avoid
the mentioned anomalies.

In this section, we regard the geometrical approaches toward studying
the thermodynamic structure of the solutions. The possible root and divergence
points of the heat capacity were obtained before, and therefore, we examine
their match/mismatch with divergences of the thermodynamical Ricci scalar. A
successful method of geometrical thermodynamics covers all characteristic
points that were observed in heat capacity. To put it more clearly, no
mismatch between divergence points of the Ricci scalar and associated points
of heat capacity should be observed and no extra divergence should exist.

The thermodynamic metrics of Weinhold \cite{WeinholdI,WeinholdII},
Ruppeiner \cite{RuppeinerI,RuppeinerII}, Quevedo
\cite{QuevedoI,QuevedoII} and HPEM \cite{HPEMI,HPEMII} are,
respectively, introduces as
\begin{equation}
ds^{2}=\left\{
\begin{array}{cc}
Mg_{ab}^{W}dX^{a}dX^{b} & Weinhold \\
&  \\
-T^{-1}Mg_{ab}^{R}dX^{a}dX^{b} & Ruppeiner \\
&  \\
\left( SM_{S}+QM_{Q}\right) \left( -M_{SS}dS^{2}+M_{QQ}dQ^{2}\right) &
Quevedo \\
&  \\
\frac{M_{S}}{M_{QQ}^{3}}\left( -M_{SS}dS^{2}+M_{QQ}dQ^{2}\right) & HPEM%
\end{array}%
\right.
\end{equation}%
where $M_{X}=\partial M/\partial X$ and $M_{XX}=\partial ^{2}M/\partial
X^{2} $. To find the divergence points of the Ricci scalar
corresponding to the mentioned metrics, we can consider the roots of Ricci
scalar's denominator. In that case, one can obtain denominators of Ricci scalar as \cite{HPEMI}
\begin{equation}
denom(\mathcal{R})=\left\{
\begin{array}{cc}
M^{2}\left( M_{SS}M_{QQ}-M_{SQ}^{2}\right) ^{2} & Weinhold \\
&  \\
M^{2}T\left( M_{SS}M_{QQ}-M_{SQ}^{2}\right) ^{2} & Ruppeiner \\
&  \\
M_{SS}^{2}M_{QQ}^{2}\left( SM_{S}+QM_{Q}\right) ^{3} & Quevedo \\
&  \\
2M_{SS}^{2}M_{S}^{3} & HPEM%
\end{array}%
\right.
\end{equation}

Regarding the finite mass, one finds some extra roots of Ricci scalar's
the denominator of all thermodynamic metrics, except HPEM case, e.g., an extra
(mismatched) divergence point appears when $M_{SS}M_{QQ}=M_{SQ}^{2}$ for
both Weinhold and Ruppeiner cases, while for Quevedo, it appears when $SM_{S}=-QM_{Q}$ or $M_{QQ}=0$. The existence of $M_{SS}$ and $M_{S}$ in
Ricci scalar's denominator of HPEM and Quevedo metrics shows that
the Ricci scalar divergences are exactly matched with bound and phase
transition points of the heat capacity.\newline

\begin{figure}[tbp]
$%
\begin{array}{cc}
\epsfxsize=6.5cm \epsffile{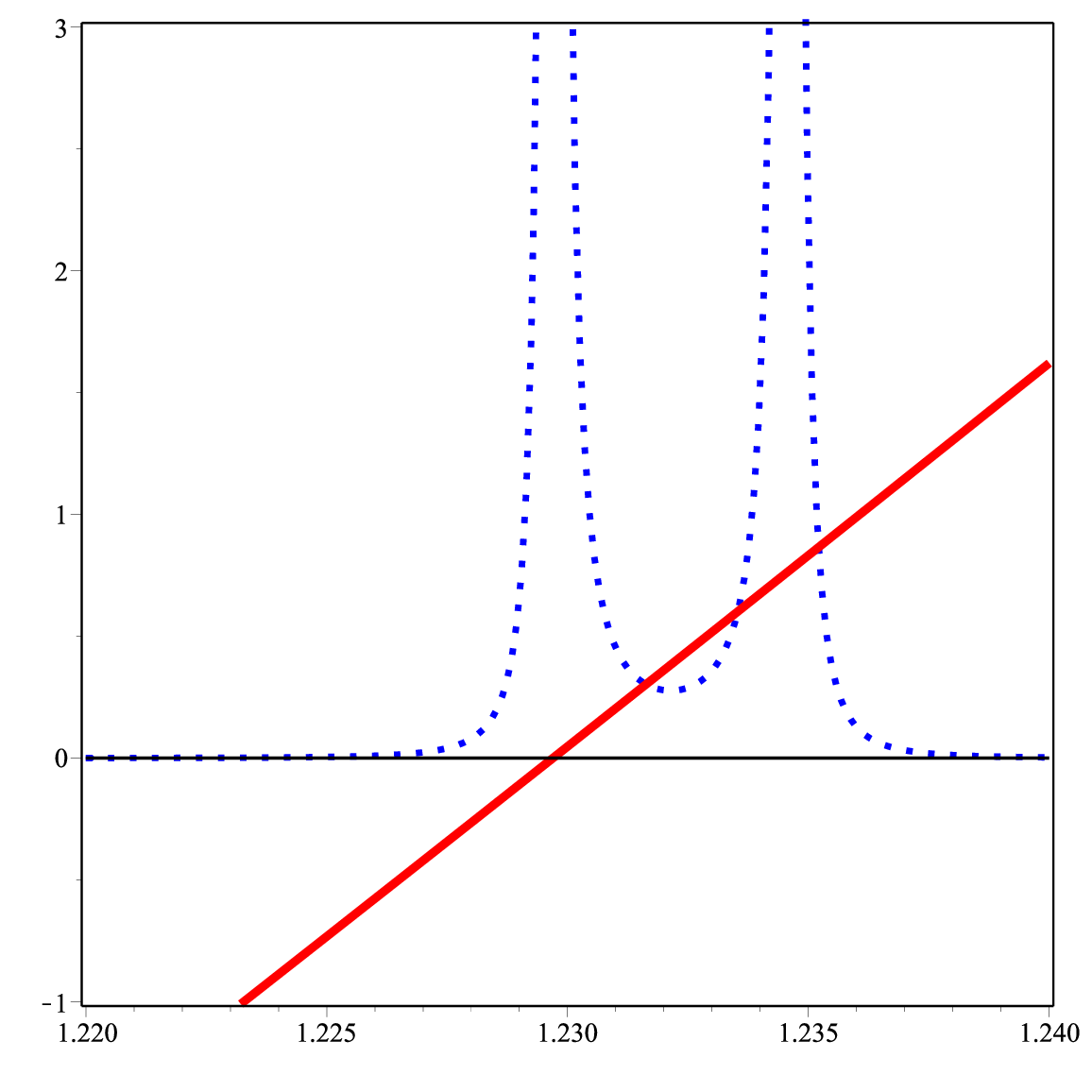} & \epsfxsize=6.5cm %
\epsffile{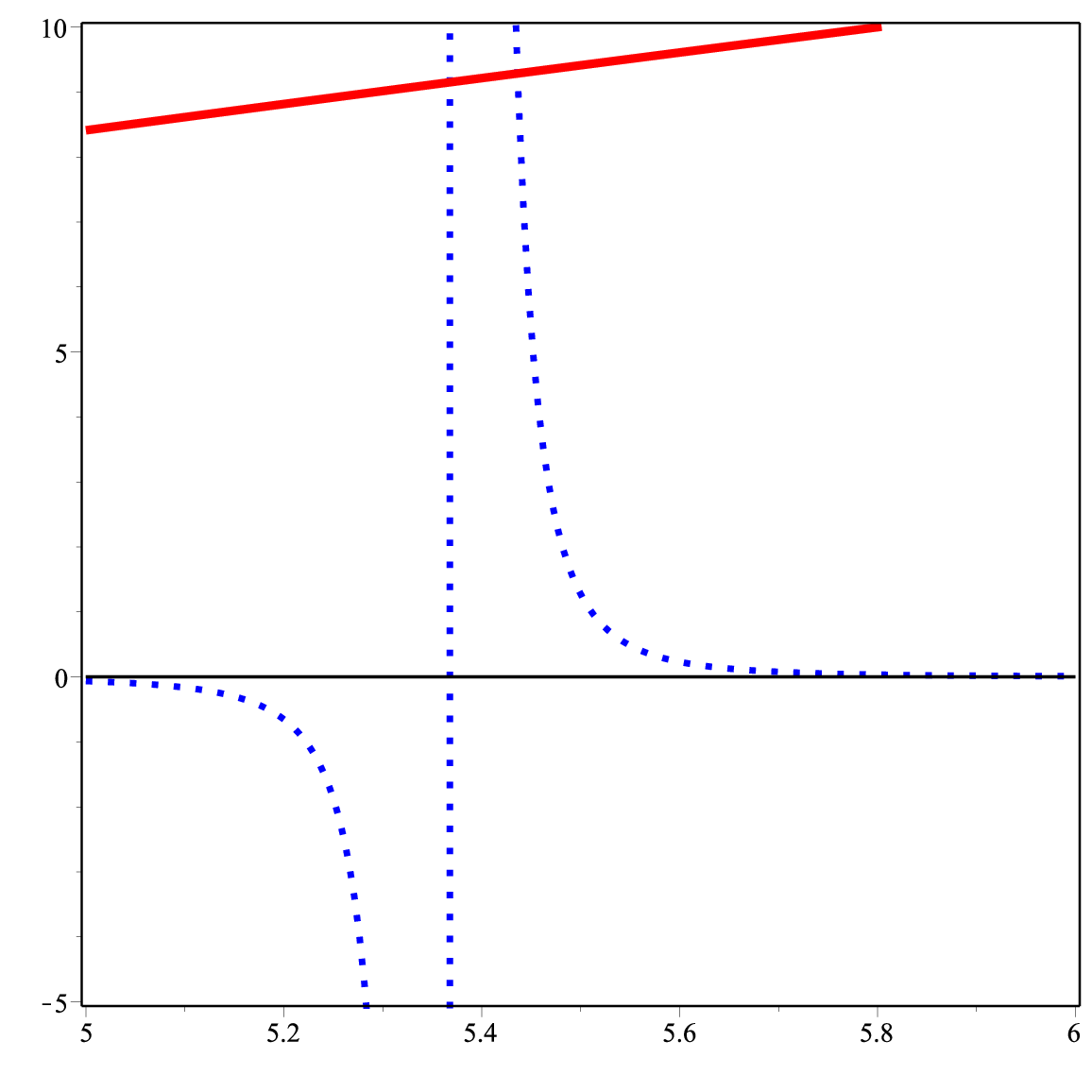}%
\end{array}
\newline
\begin{array}{c}
\hspace{3cm} \epsfxsize=6.5cm \epsffile{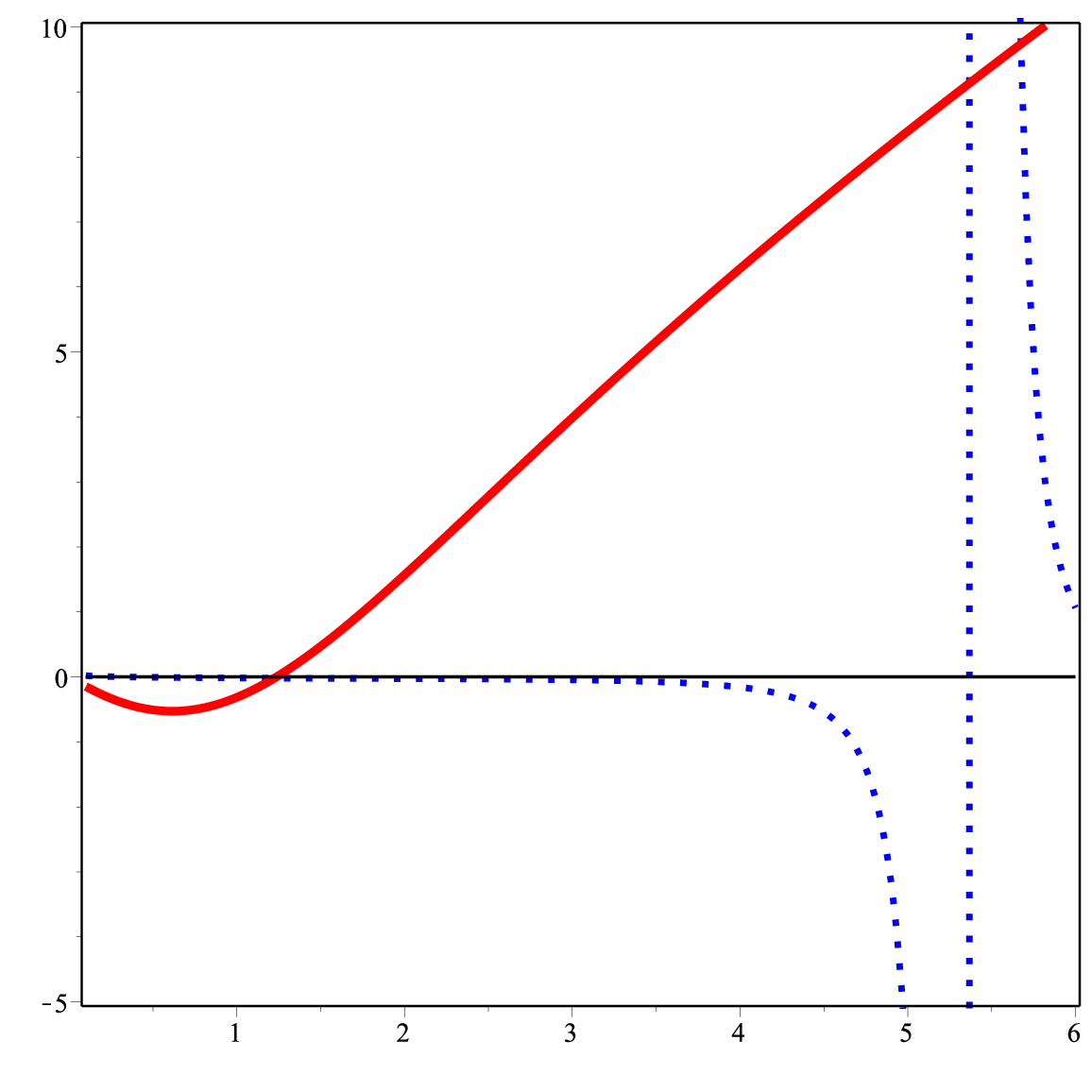}%
\end{array}
\newline
\begin{array}{cc}
\epsfxsize=6.5cm \epsffile{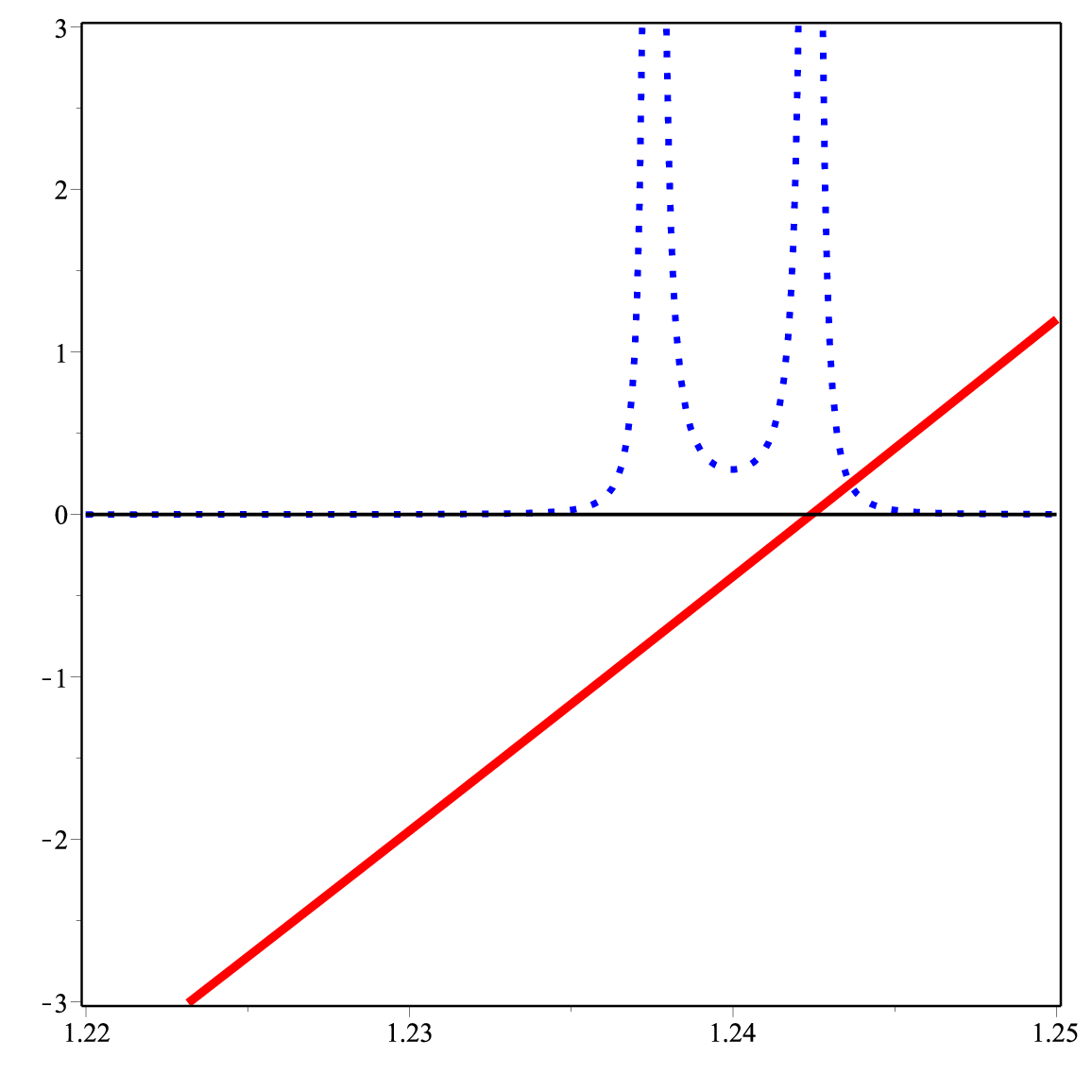} & \epsfxsize=6.5cm %
\epsffile{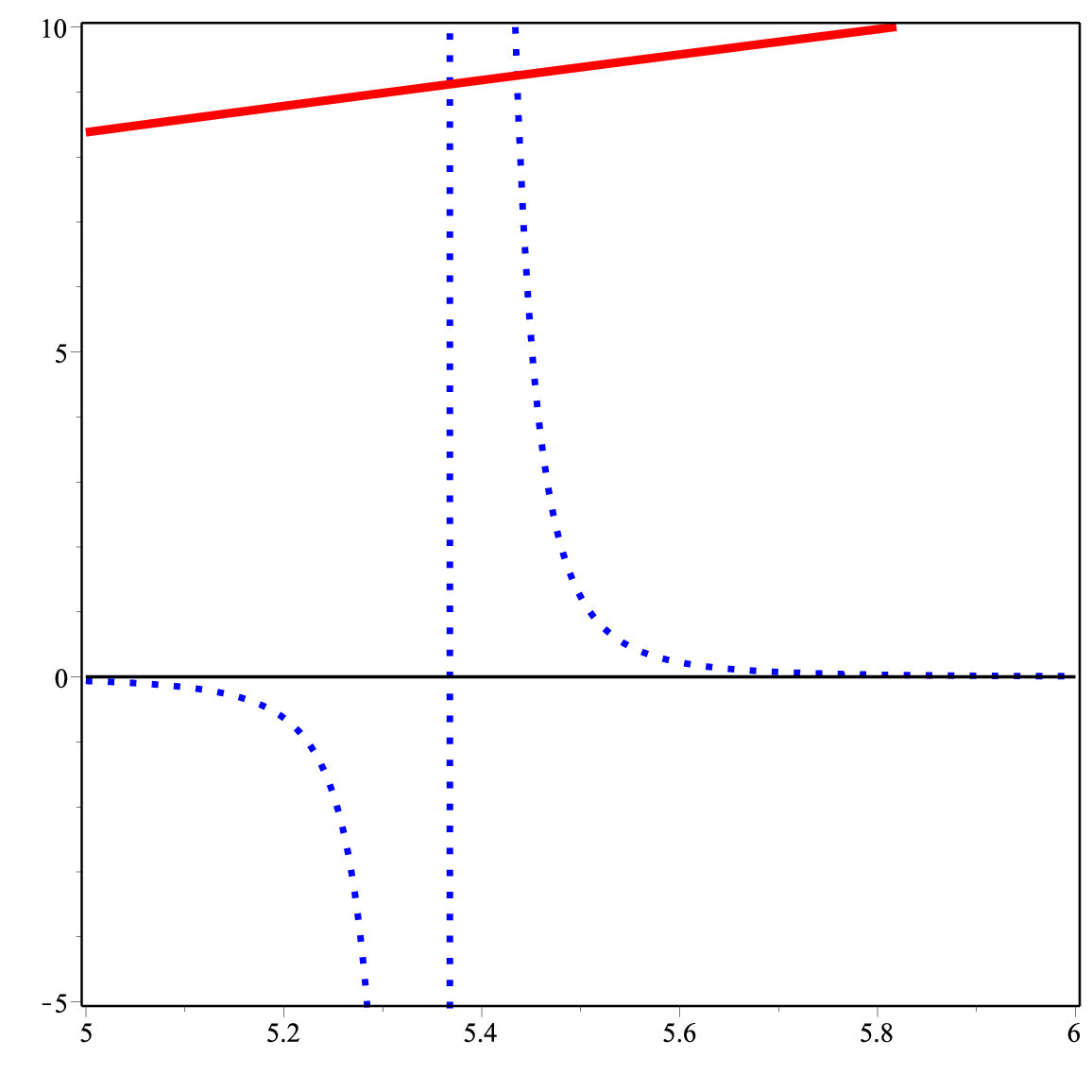}%
\end{array}
$%
\caption{Weinhold's metric: $C_{Q}$ (continuous line) and $\mathcal{R}$
(dotted line) versus $r_{+}$ for $Q=c=c_{1}=1$, $\Lambda=-1$, $l=0.01$ and $%
m=2$. \newline
Upper panels: $\protect\alpha=-1$, middle panel: $\protect\alpha=0$ and
lower panels: $\protect\alpha=+1$.}
\label{FigW}
\end{figure}

\begin{figure}[tbp]
$%
\begin{array}{cc}
\epsfxsize=6.5cm \epsffile{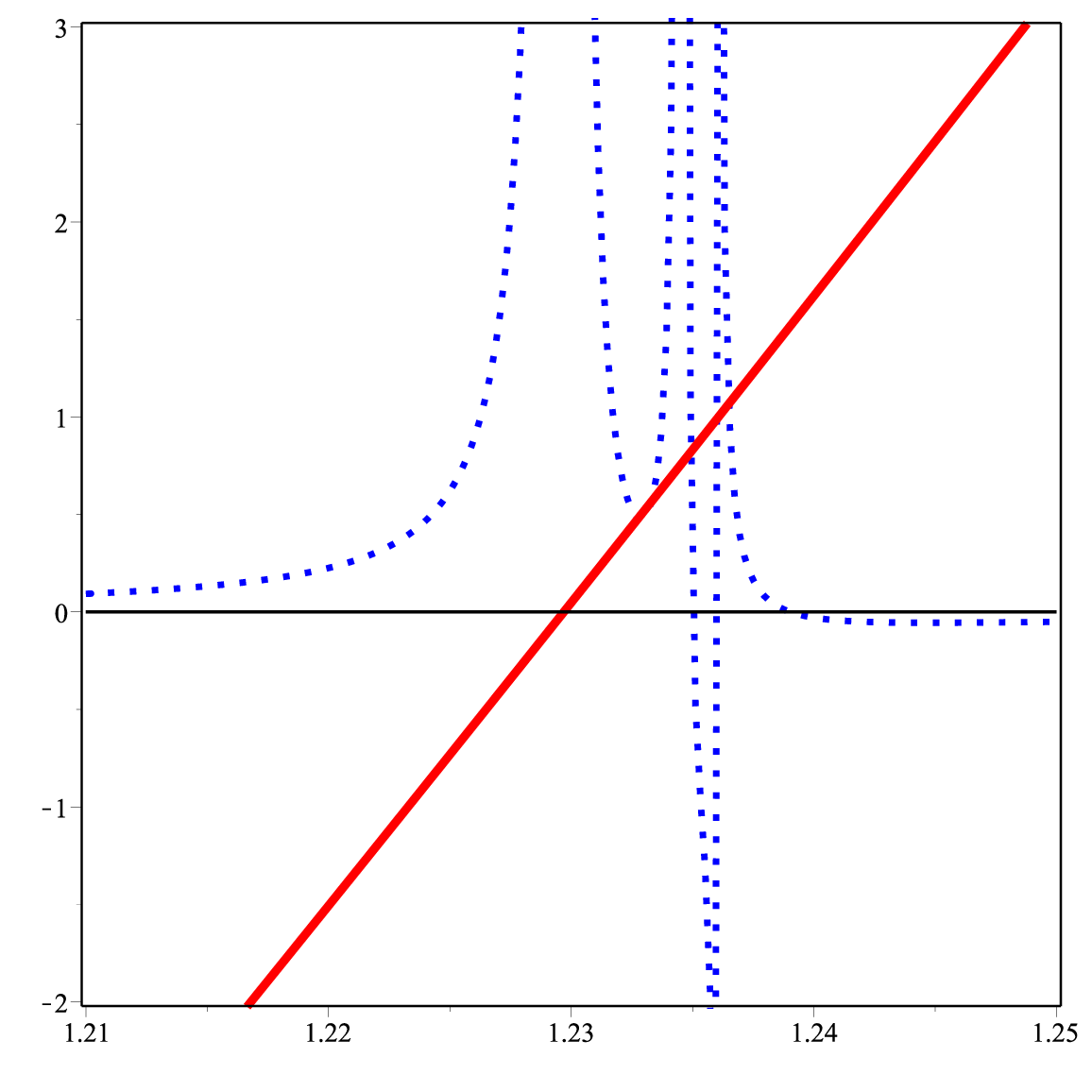} & \epsfxsize=6.5cm %
\epsffile{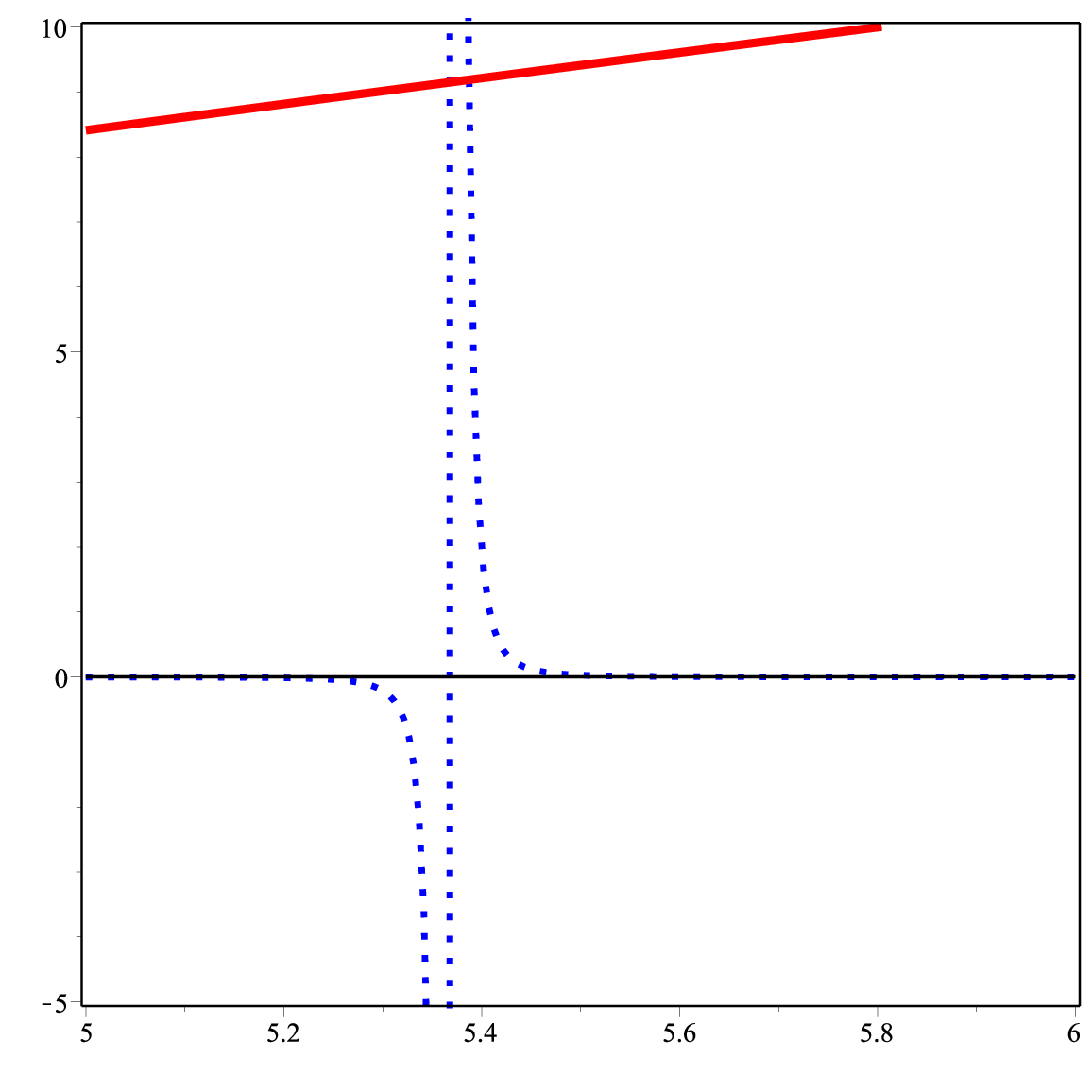}%
\end{array}
\newline
\begin{array}{c}
\hspace{3cm} \epsfxsize=6.5cm \epsffile{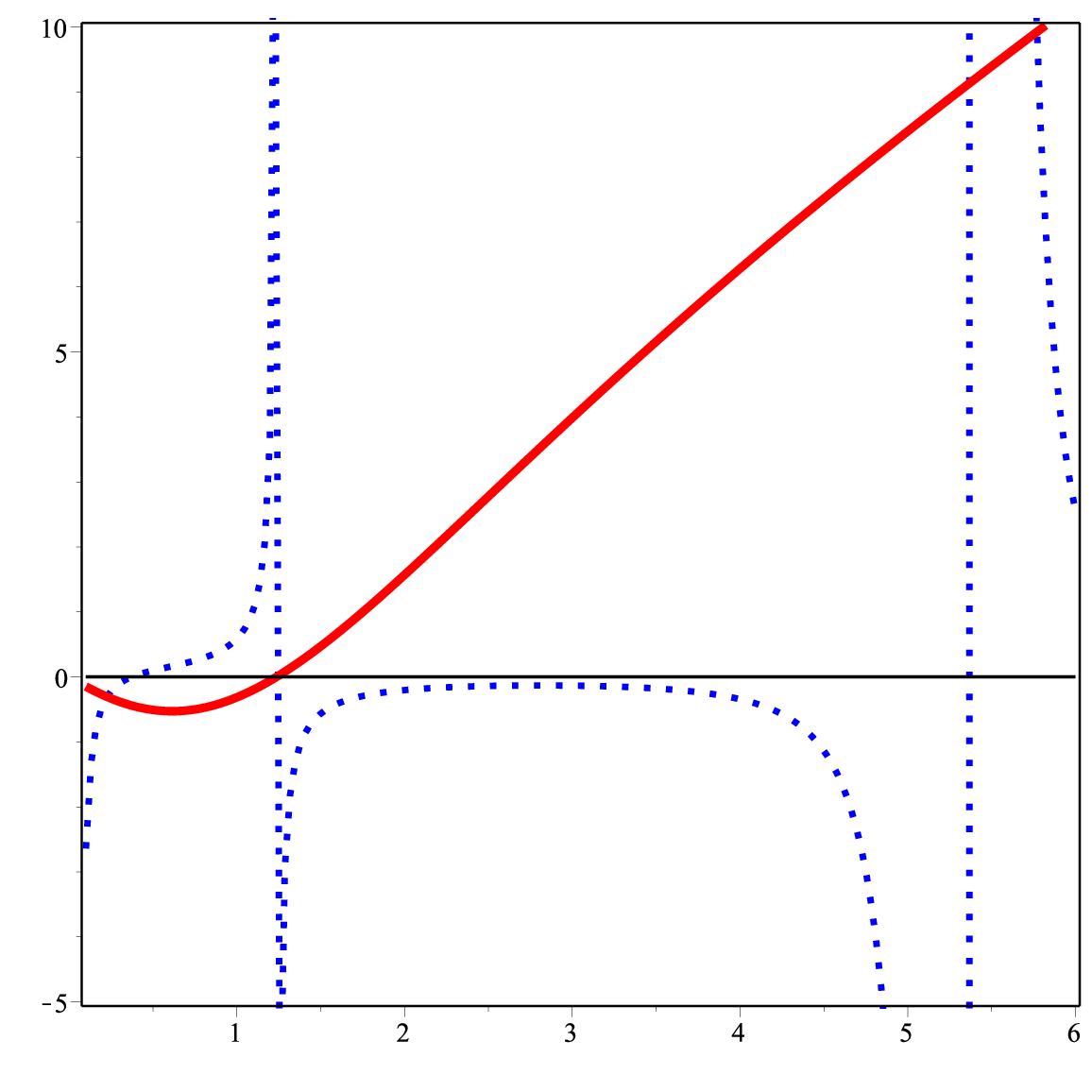}%
\end{array}
\newline
\begin{array}{cc}
\epsfxsize=6.5cm \epsffile{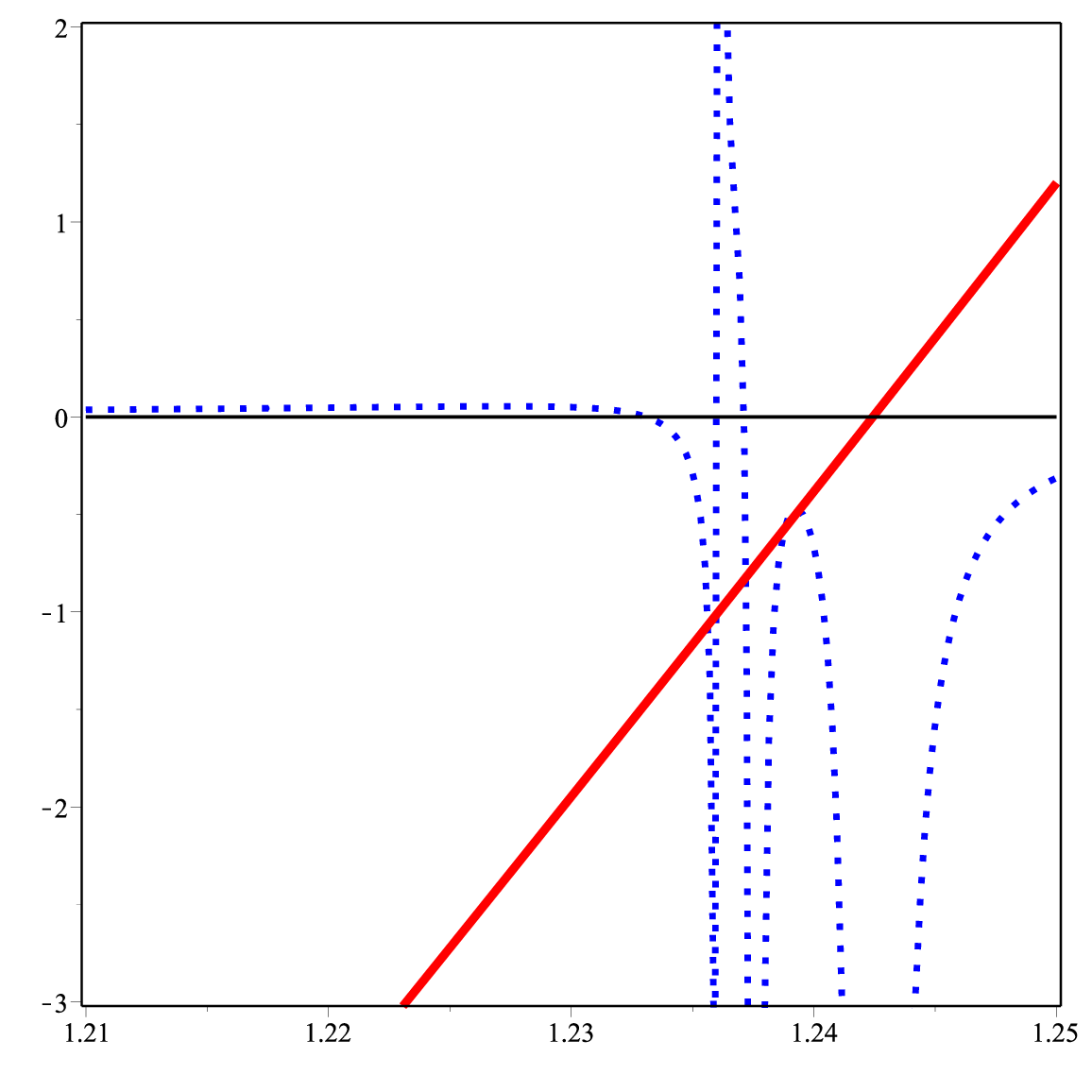} & \epsfxsize=6.5cm %
\epsffile{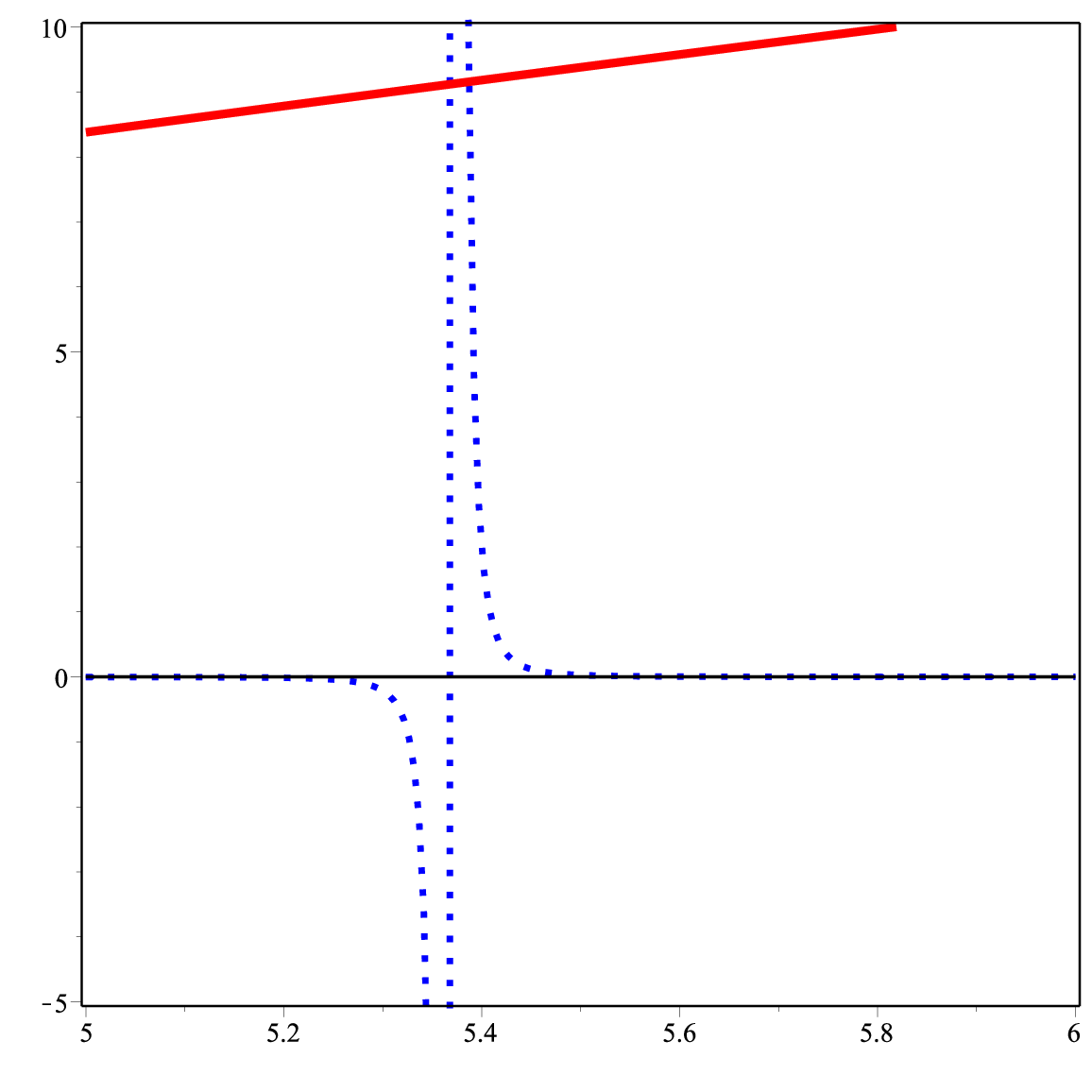}%
\end{array}
$%
\caption{Ruppeiner's metric: $C_{Q}$ (continuous line) and $\mathcal{R}$
(dotted line) versus $r_{+}$ for $Q=c=c_{1}=1$, $\Lambda=-1$, $l=0.01$ and $%
m=2$. \newline
Upper panels: $\protect\alpha=-1$, middle panel: $\protect\alpha=0$ and
lower panels: $\protect\alpha=+1$.}
\label{FigR}
\end{figure}
\begin{figure}[tbp]
$%
\begin{array}{cc}
\epsfxsize=6.5cm \epsffile{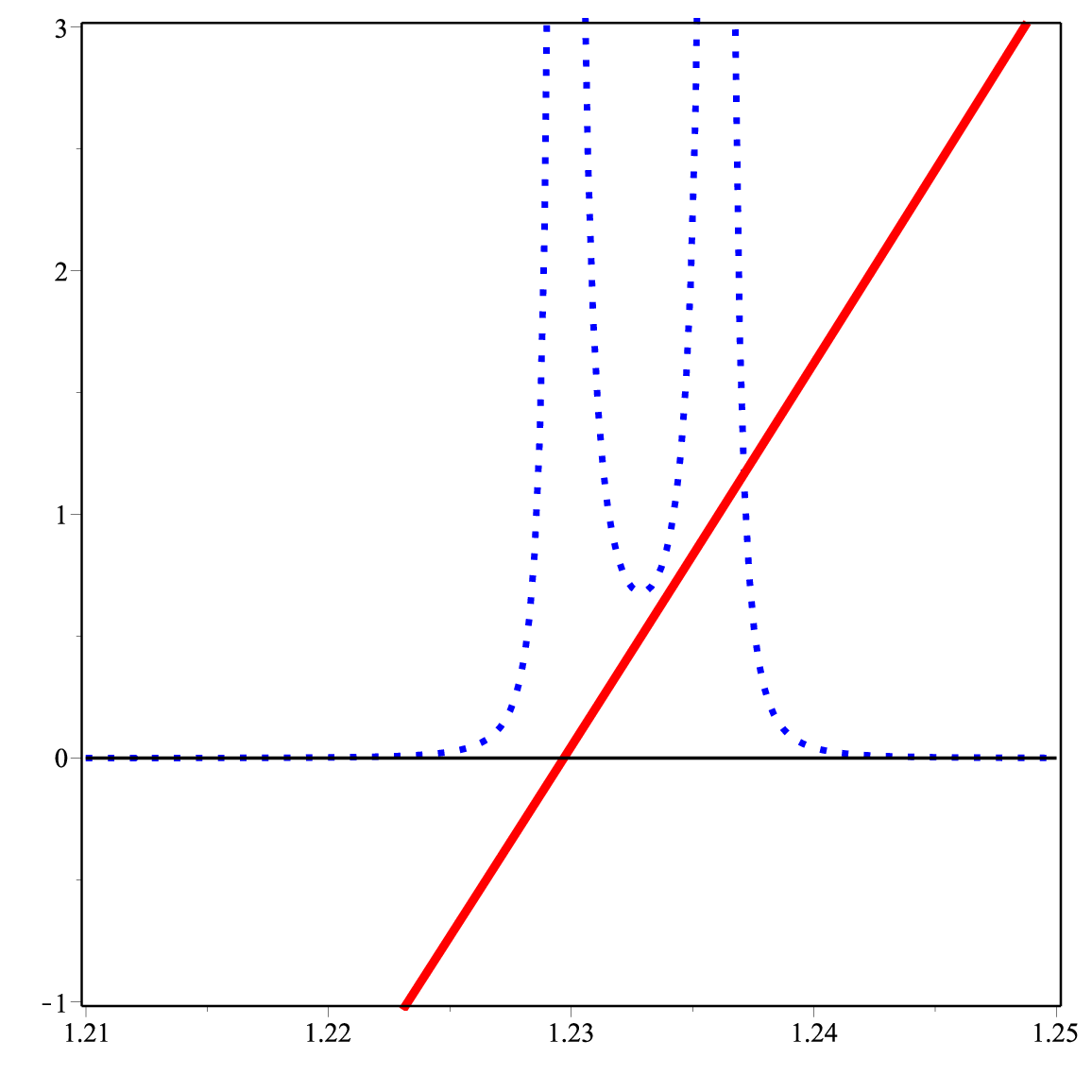} & \epsfxsize=6.5cm %
\epsffile{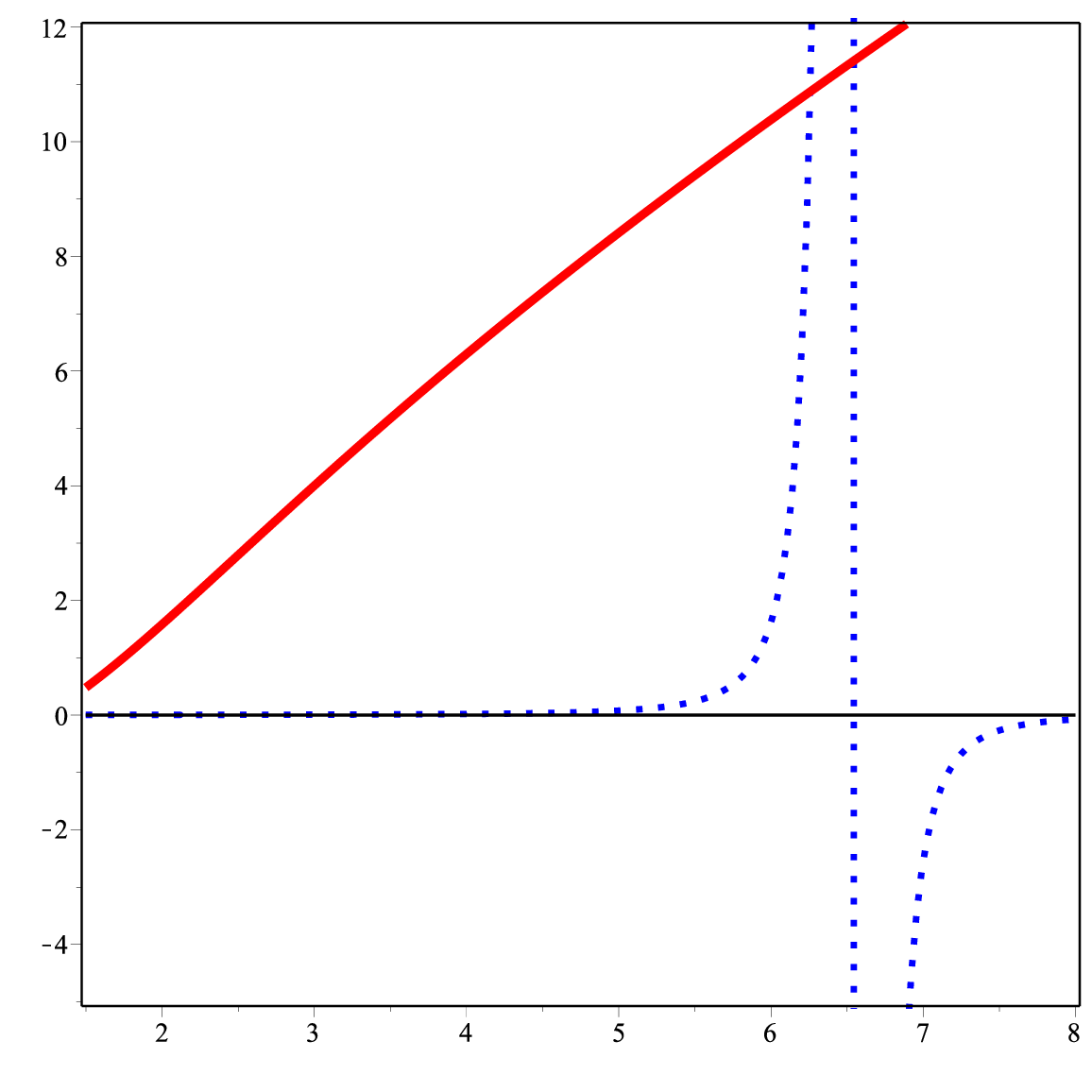}%
\end{array}
\newline
\begin{array}{c}
\hspace{3cm} \epsfxsize=6.5cm \epsffile{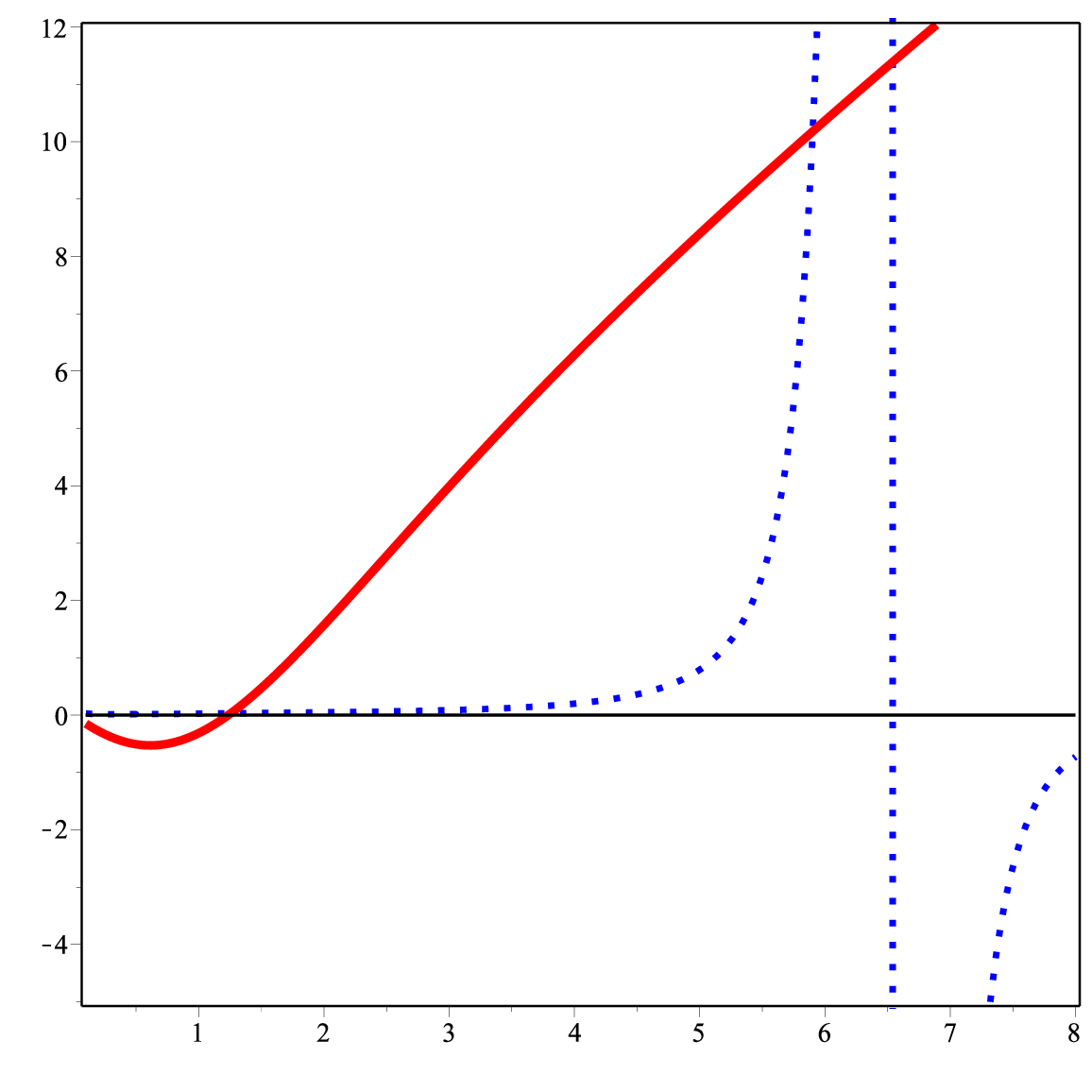}%
\end{array}
\newline
\begin{array}{cc}
\epsfxsize=6.5cm \epsffile{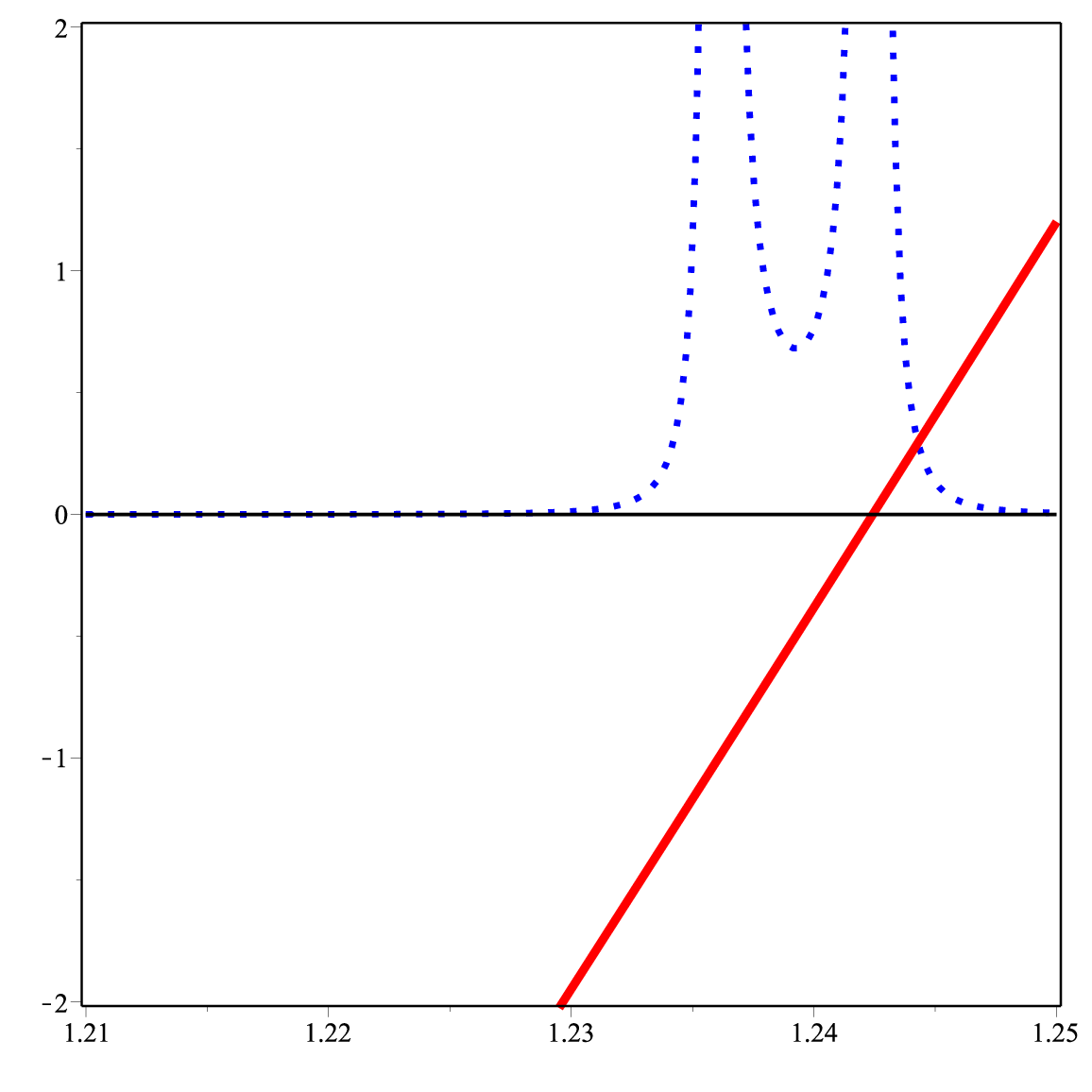} & \epsfxsize=6.5cm %
\epsffile{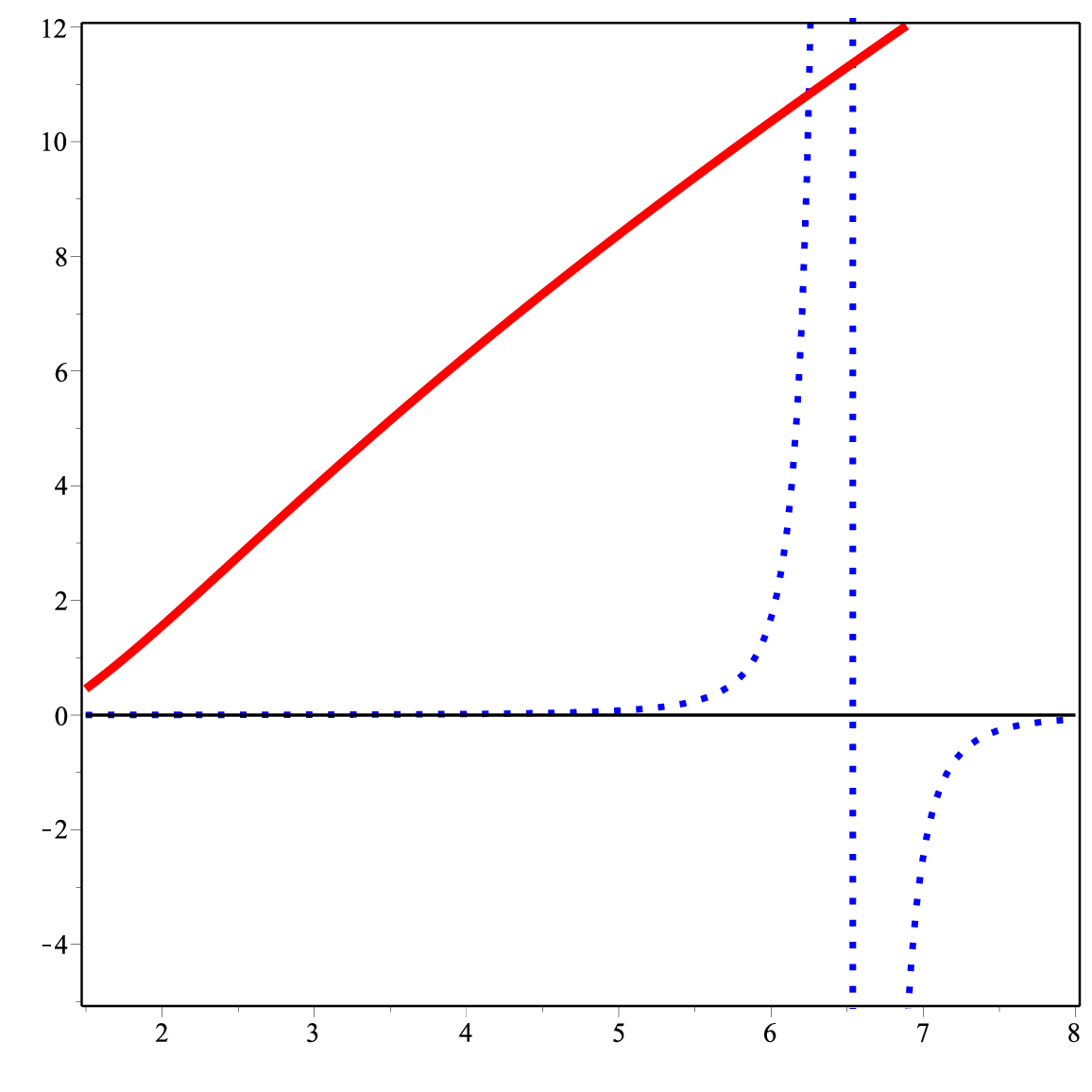}%
\end{array}
$%
\caption{Quevedo's metric: $C_{Q}$ (continuous line) and $\mathcal{R}$
(dotted line) versus $r_{+}$ for $Q=c=c_{1}=1$, $\Lambda=-1$, $l=0.01$ and $%
m=2$. \newline
Upper panels: $\protect\alpha=-1$, middle panel: $\protect\alpha=0$ and
lower panels: $\protect\alpha=+1$.}
\label{FigQ}
\end{figure}
\begin{figure}[tbp]
$%
\begin{array}{cc}
\epsfxsize=6.5cm \epsffile{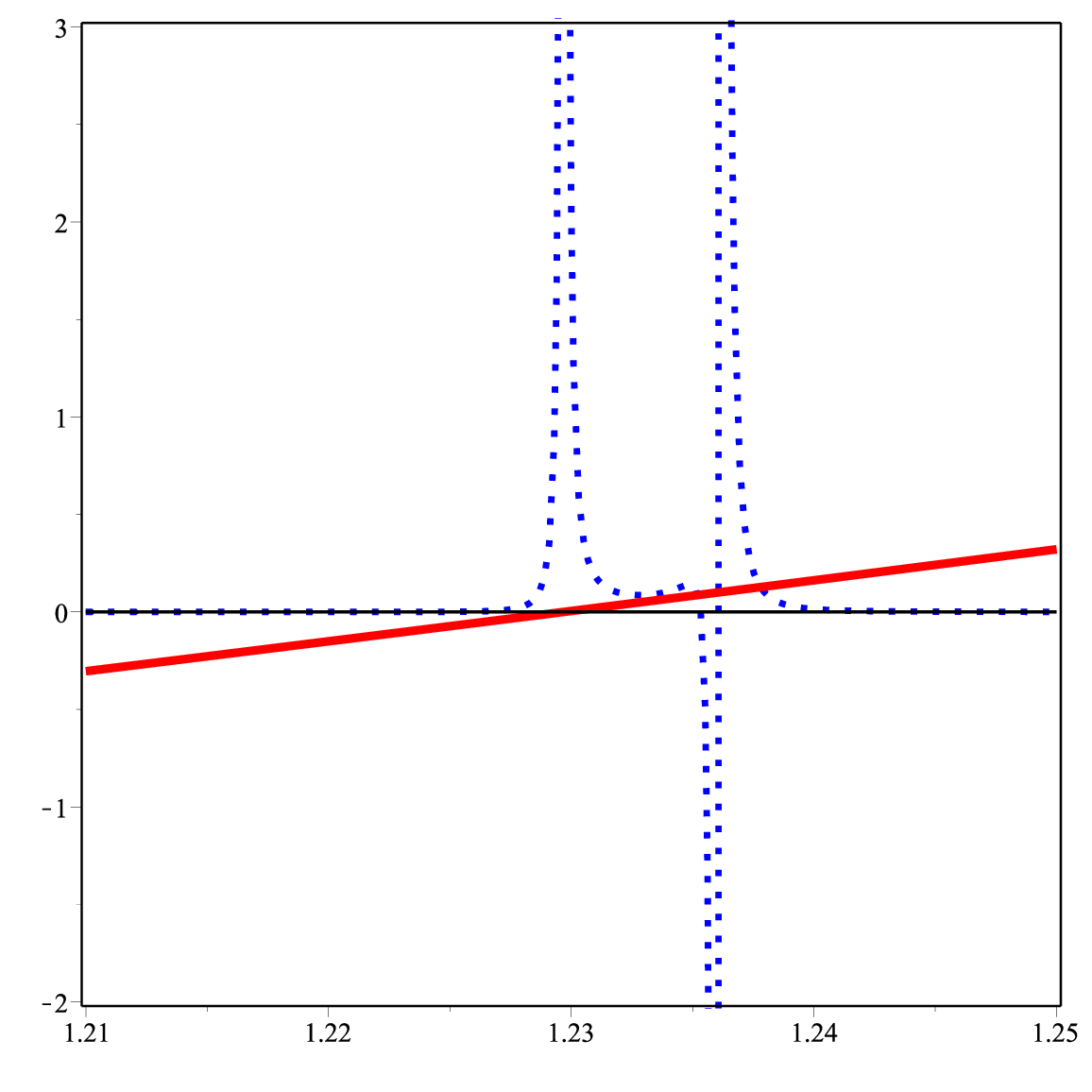} & \epsfxsize=6.5cm %
\epsffile{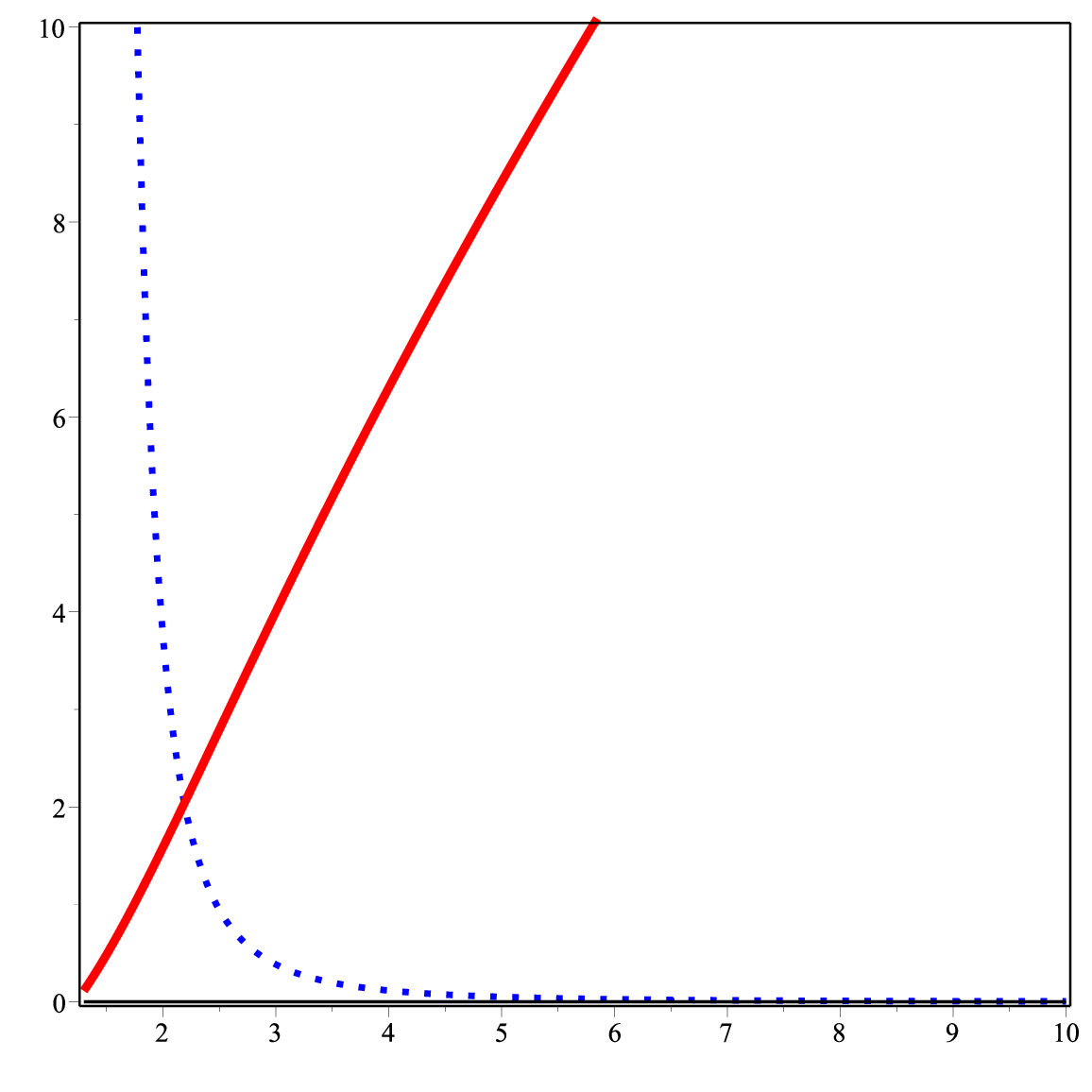}%
\end{array}
\newline
\begin{array}{c}
\hspace{3cm} \epsfxsize=6.5cm \epsffile{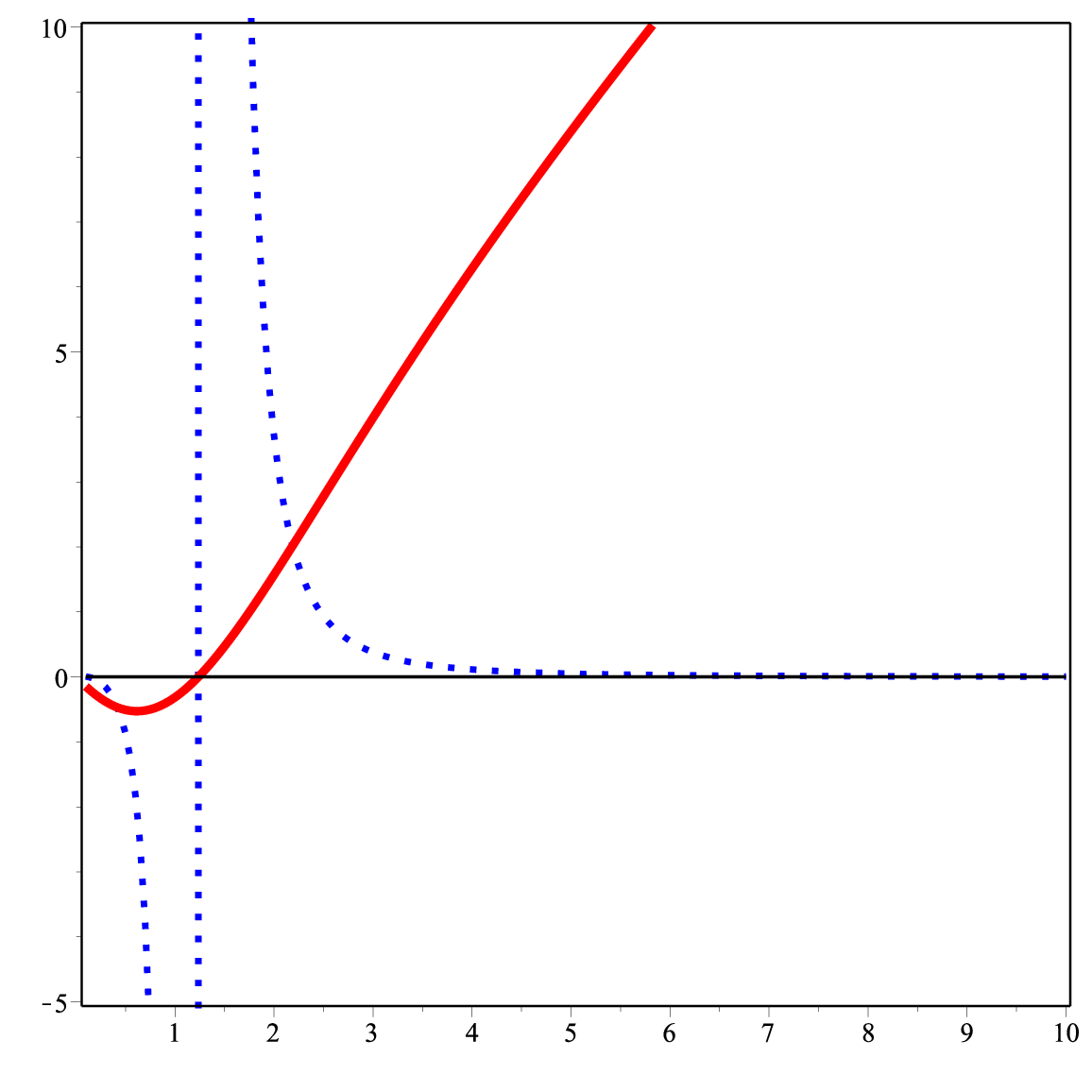}%
\end{array}
\newline
\begin{array}{cc}
\epsfxsize=6.5cm \epsffile{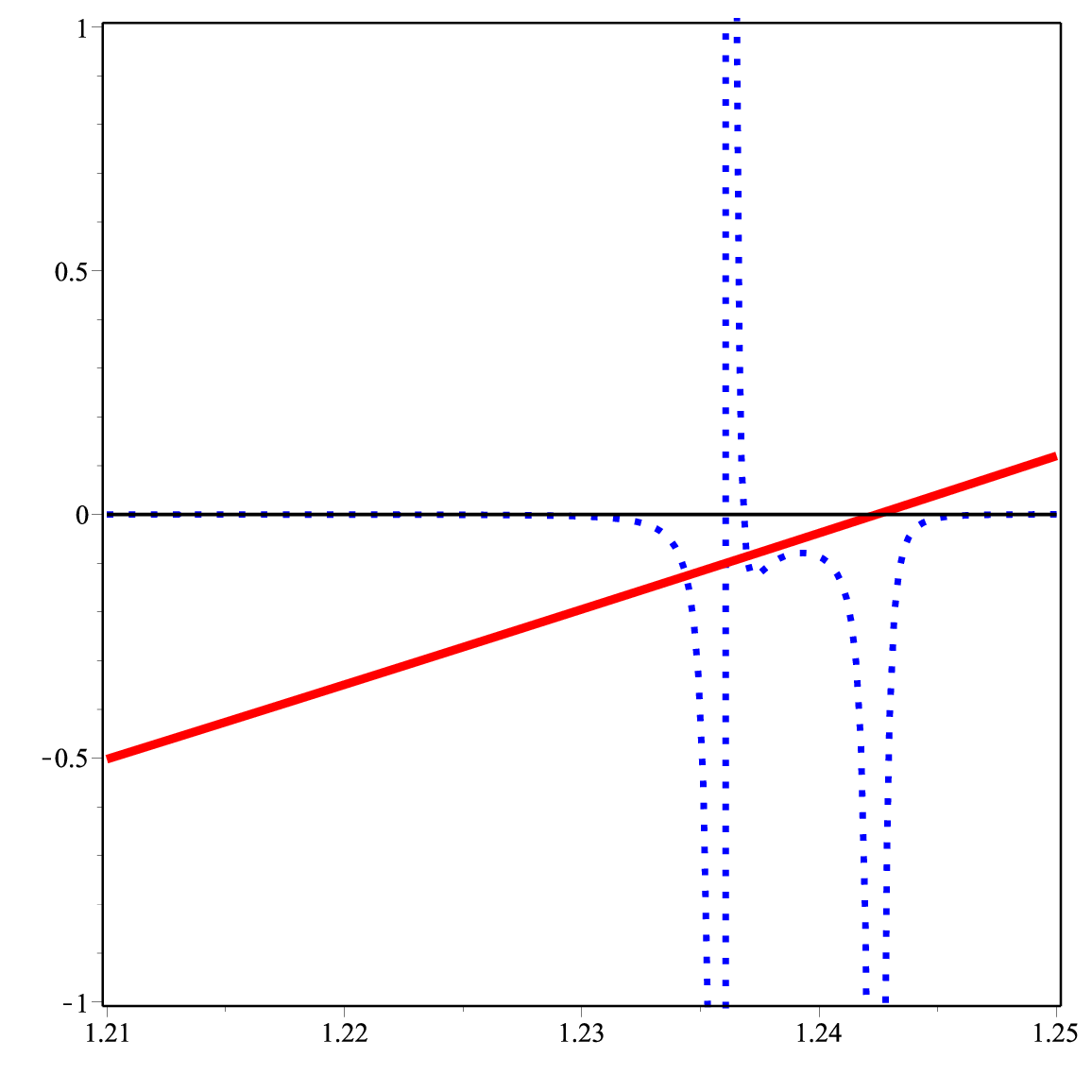} & \epsfxsize=6.5cm %
\epsffile{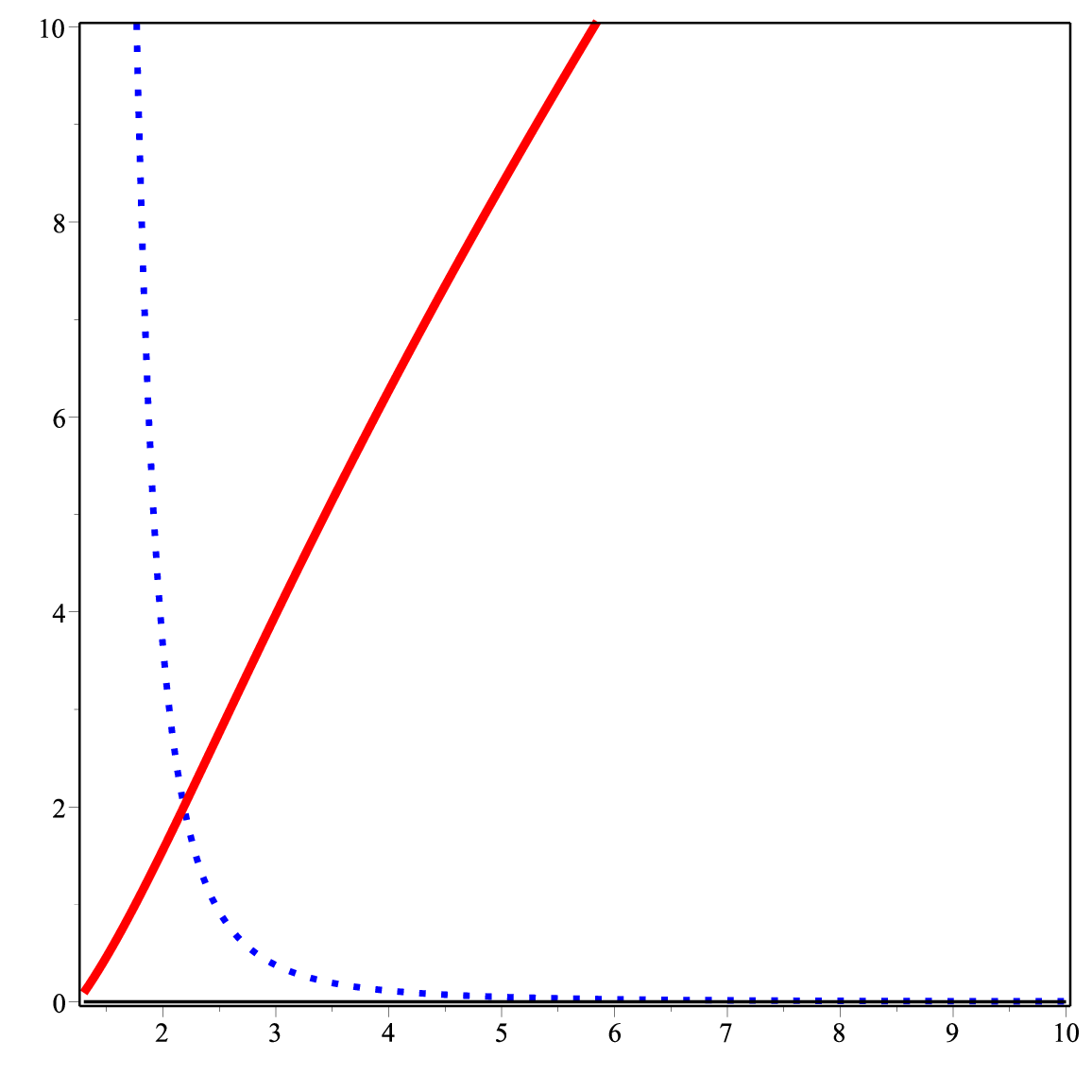}%
\end{array}
$%
\caption{HPEM's metric: $C_{Q}$ (continuous line) and $\mathcal{R}$ (dotted
line) versus $r_{+}$ for $Q=c=c_{1}=1$, $\Lambda=-1$, $l=0.01$ and $m=2$.
\newline
Upper panels: $\protect\alpha=-1$, middle panel: $\protect\alpha=0$ and
lower panels: $\protect\alpha=+1$.}
\label{FigN}
\end{figure}

Since the analytical discussion is impossible, we use numerical
calculations and give the results in some diagrams (see Figs. \ref{FigW}, %
\ref{FigR}, \ref{FigQ} and \ref{FigN}).

Comparing the plot of the Ricci scalar and heat capacity, we find that for
zero $\alpha$, all the Ricci scalars have an extra divergence far from the
root of heat capacity, except HPEM one. In other words, the heat capacity is
a finite nonzero function in such an extra divergence. So, except for the
HPEM case, other thermodynamic metrics have the mentioned anomaly.

For the case of nonzero $\alpha$, the same behavior is seen for
the mentioned cases far from the root of heat capacity. Also, one may observe some extra divergences near the root of
heat capacity. The existence of a divergence at the root of heat
capacity is expected to characterize the bound point. However,
there may exist additional divergence in this region, which is due
to the logarithmic form of thermal fluctuations of entropy. So, for
nonzero $\alpha$, we should take care of the mentioned extra
divergence of the Ricci scalar, which is due to the functional form
of corrected entropy. As a final point, we should note that all
the mentioned figures and discussions in this section are related
to the linear Maxwell solution and for the nonlinear one, the same
behaviour is observed, which is ignored for the sake of brevity.

\newpage

\section{Conclusions and discussions} \label{sec5}

It is believed that the logarithmic correction to the entropy of a
black object is obtained in almost all approaches to quantum gravity. Therefore, one may consider it as a universal model-independent result. So,
the first step is to calculate the non-perturbative entropy and temperature
with the conventional method. As the second step, we should build the argument
of logarithmic function which is proportional to "$S_{0}\times T^2$". Since
dimensions of temperature ($T$) and entropy ($S_{0}$) in arbitrary $d$%
-dimensions are, respectively, $L^{-1}$ and $L^{d-2}$, one finds that the
argument of logarithm ($S_{0}T^2$) is dimensionless only in 4-dimensions.
So, we have to define dimensionless $T$ and $S_{0}$ in arbitrary dimensions $%
d \neq 4$. The third step is devoted to the calculation of the corrected entropy as presented
in (\ref{Corrected-Entropy}). As the last step, we can obtain different
quantities related to the stability of the solutions (such as heat capacity and
free energy) with new corrected entropy to find the stability criteria in
the presence of thermal fluctuations.\newline
Now, we summarize our results here. First, we have developed the corrected
thermodynamical equation of states for static charged BTZ black hole with
AdS asymptotes in the context of massive gravity by considering small
statistical fluctuations to the system. In particular, we derived the
dimensionless internal energy and heat capacity. In case when thermal
fluctuation is considered, we must have nonzero $\alpha$. Remarkably, the
negative $\alpha$ does not impact the stability of the AdS solution. However,
for positive $\alpha$ the situation is not trivial stability can be
achieved under certain restrictions.

We have found that the specific heat is negative for the cases of $\alpha =
0 $ and $\alpha = \pm 1$, hence the black hole is unstable, and there exists
a singular point that is not a phase transition point because the black
hole is unstable before and after that point. Therefore, the final stage of the black hole stability depends on the correction coefficient and non-linear parameter. Hence the mentioned black remnants in the introduction section which was in the absence of the logarithmic correction may be happen again which is the subject of the future work.

In the presence of thermal
fluctuations, the Helmholtz free energy has a maximum which may be a sign of
an instability or phase transition.

Furthermore, we extended our study for the case of nonlinearly charged AdS
black hole. Here, we emphasized that there is not any considerable change in
the behavior of Gibbs free energy under thermal correction. For parameter $%
\beta =0.5$, we can see linear behavior of specific heat with a completely positive value for $\alpha =-1$. 
In presence of the thermal fluctuations with $\alpha =-1$, we
found large/small phase transition at $r_{+}\approx 0.4$ through a first order phase transition. For $\alpha =1$%
, there is a minimum of specific heat at the negative region and therefore,
the specific heat has two zeros for this case. We found that the critical points and Van
der Waals-like behavior don't occur here.

According to the results of Ref. \cite{1201.0463}, dynamical
stability of a black hole is equivalent to its thermodynamic
stability. Such equality may break in presence of thermal
fluctuation. Ignoring thermal fluctuations (classical
description), the nonlinearly charged AdS black hole in massive
gravity may be stable thermodynamically and hence dynamically. But
in presence of thermal fluctuations it is unstable
thermodynamically at small sizes, but it is stable dynamically. It
may be a classical equivalent which violated in presence of
thermal fluctuations. It means that log correction has no any
effect on dynamical stability.

In order to investigate the thermodynamic structure, we utilized
geometrical thermodynamics. The numerical calculations show that
in thermal equilibrium, all the Ricci scalars have an extra
divergence far from the root of heat capacity, except HPEM one. In
the case of a thermally disturbed system, the same behavior is
seen for the mentioned cases far from the root of heat capacity.
Besides, one may observe some extra divergences near the root of
heat capacity. The existence of a divergence at the root of heat
capacity is expected to characterize the bound point. However,
there may exist additional divergence in this region which is due
to the logarithmic form of thermal fluctuations of entropy.

In addition to what was mentioned above, it is interesting to use
the results of this paper to study the black hole remnants.
Besides, studying the logarithmic corrections to the entropy may
be considered via microscopic degrees of freedom of black objects
in the context of conformal field theory. Moreover, it may be
useful to focus on the information loss paradox based on the
obtained results. Furthermore, it is interesting to investigate
the shear viscosity to entropy ratio of this model due to the
mentioned logarithmic correction and examine its lower bound.
Finally, it will be worthy to consider higher order correction terms
\cite{2007.15401} to investigate black hole thermodynamics in the case of various modifications of gravity \cite{CANTATA:2021ktz,Gonzalez:2011dr,Capozziello:2012zj}. We
leave these issues for future work.

\section*{Acknowledgements}
{We are thankful to the Editor and the anonymous Referees for their valuable recommendations and feedback. \.{I}.S. wishes to extend gratitude for the networking support received from COST Action CA18108, dedicated to quantum gravity phenomenology in the multi-messenger approach. Additionally, he appreciates the generous support from T\"{U}B\.{I}TAK, ANKOS, and SCOAP3.}


\begin{thebibliography}{99}
\bibitem{BasuPRD2016} S. Basu, J. Bhattacharyya, D. Mattingly, M. Roberson, Phys. Rev. D 93, 064072 (2016).
\bibitem{JaniszewskiJHEP2013} S. Janiszewski and A. Karch, JHEP 02, 123 (2013).
\bibitem{GriffinPRL2013} T. Griffin, P. Horava and C. M. Melby-Thompson, Phys. Rev. Lett. 110,  081602 (2013).
\bibitem{Carlip1995} S. Carlip, Class. Quantum Gravit. 12, 2853 (1995).
\bibitem{Ashtekar2002} A. Ashtekar, J. Wisniewski and O. Dreyer, Adv. Theor. Math. Phys. 6, 507 (2002).
\bibitem{Sarkar2006} T. Sarkar, G. Sengupta and B. N. Tiwari, JHEP 11, 015 (2006).
\bibitem{Witten1998} E. Witten, Adv. Theor. Math. Phys. 2, 505 (1998).
\bibitem{Carlip2005} S. Carlip, Class. Quantum Gravit. 22, R85 (2005).

\bibitem{Newmassive} E. A. Bergshoeff, O. Hohm and P. K. Townsend, Phys. Rev. Lett. 102, 201301 (2009).

{\bibitem{deRham:2010kj}
C.~de Rham, G.~Gabadadze and A.~J.~Tolley,
Phys. Rev. Lett.  {106}, 231101 (2011).
\bibitem{Kanzi:2020cyv}
S.~Kanzi, S.~H.~Mazharimousavi and \.I.~Sakall\i{},
Annals Phys. {422}, 168301 (2020).}

\bibitem{Vegh} D. Vegh, [arXiv:1301.0537].

{\bibitem{Chen:2017dsy}
C.~F.~Chen and A.~Lucas,
Phys. Lett. B  {774}, 569-574 (2017).}

\bibitem{Zhang2016} H. Zhang and X. Z. Li, Phys. Rev. D 93, 124039 (2016).
\bibitem{Hu2016} Y. P. Hu, H. F. Li, H. B. Zeng and H. Q. Zhang, Phys. Rev. D 93, 104009 (2016).
\bibitem{Zeng2016} X. X. Zeng, H. Zhang and L. F. Li, Phys. Lett. B 756, 170 (2016).
{\bibitem{ras}D. A. Rasheed, hep-th/9702087 [hep-th].
\bibitem{ras1}A. Chamblin, R. Emparan, C. V. Johnson and R. C. Myers, 	Phys. Rev. D 60, 104026 (1999).
\bibitem{ras2}R.-G. Cai and A. Wang, Phys. Rev. D 70, 064013 (2004).
\bibitem{ras3}H. A. Gonzalez, M. Hassaine and C. Martinez, Phys. Rev. D 80 (2009) 104008.
\bibitem{ras4} O. Miskovic and R. Olea, Phys. Rev. D 77, 124048 (2008).
\bibitem{ras5}D. P. Sorokin, Fortsch.Phys. 70 (2022)  2200092.
\bibitem{ras6}H. Babaei-Aghbolagh, K. B. Velni, D. M. Yekta and H. Mohammadzadeh,  Phys. Lett. B 829 (2022) 137079.
\bibitem{ras7}K. Lechner, P. Marchetti, A. Sainaghi and D. P. Sorokin, Phys.Rev.D 106 (2022)   016009.
\bibitem{ras8} G. Arenas-Henriquez, F. Diaz and Y. Novoa, JHEP 05 (2023) 072.


\bibitem{Born:1933pep}
M.~Born and L.~Infeld,
Nature  {132},  1004  (1933).

\bibitem{Born:1934gh}
M.~Born and L.~Infeld,
Proc. Roy. Soc. Lond. A {144},   425  (1934).

\bibitem{Dirac:1962iy}
P.~A.~M.~Dirac,
Proc. Roy. Soc. Lond. A  {268}, 57 (1962).
\bibitem{ter} A. Belin, L.-Yan Hung, A. Maloney, S. Matsuura, R. C. Myers and T. Sierens, JHEP 12 (2013) 059.
\bibitem{ter1}W. Cong, D. Kubiznak, R. Mann and M. Visser, JHEP 08 (2022) 174.
\bibitem{ter2}P. Bueno, P. A. Cano, J. Moreno and G. v. d. Velde, Phys.Rev.D 107 (2023)   064050.
\bibitem{ter3}B. Eslam Panah, Kh. Jafarzade and A. Rincon, arXiv:2201.13211.


}

\bibitem{1} S. Das, P. Majumdar, R.K. Bhaduri,   Class. Quant. Grav. 19, 2355 (2002).

\bibitem{Addazi:2021xuf}
A.~Addazi,   \textit{et al.}  
Prog. Part. Nucl. Phys. {125}, 103948 (2022).
 


\bibitem{2} S. S. More,  
Class. Quant. Grav. 22, 4129 (2005).


\bibitem{o} S. Hawking and D. N. Page, Commun. Math. Phys. 87, 577 (1983).

\bibitem{01ab} G. W. Gibbons, S. W. Hawking and M. J. Perry, Nucl. Phys. B
138, 141 (1978).

\bibitem{hawk} G. W. Gibbons and S. W. Hawking, Phys. Rev. D. 15, 2752
(1977).

\bibitem{hawkis} J. Meixner,
Arch. Ration. Mech. Anal. {57}, 281 (1975).

\bibitem{Sakalli:2022xrb} \.I.~Sakalli and S.~Kanzi,
Turk. J. Phys.  {46},   51 (2022).

\bibitem{4} J. Sadeghi, B. Pourhassan, M. Rostami,  Phys. Rev. D 94, 064006
(2016).

\bibitem{5} R. K. Kaul, P. Majumdar,  Phys. Rev. Lett. 84, 5255 (2000).

\bibitem{7} S. Carlip, Class. Quant. Grav. 17, 4175 (2000).

\bibitem{8} A. Sen,  Gen. Rel. Grav. 44, 1947 (2012).

\bibitem{9} A. Sen, JHEP 1304, 156
(2013).

\bibitem{10} N. Morales-Dur\'{a}n, et al.,  Eur. Phys. J. C76, 559 (2016).

\bibitem{new2} D. Grumiller, A. Perez, D. Tempo, R. Troncoso, 
	JHEP08 (2017) 107.

\bibitem{11} A. Pathak, A. P. Porfyriadis, A. Strominger, O. Varela,
 JHEP 1704, 090 (2017).

\bibitem{12} B. Pourhassan, M. Faizal,   EPL 111, 40006 (2015).

\bibitem{13} M. Faizal, B. Pourhassan,   Phys. Lett. B 751, 487 (2015).


\bibitem{15} F. Hammad, M. Faizal,   Int. J. Mod. Phys. D 25, 1650080 (2016).


\bibitem{17} C. Gao, Y-G. Shen,  Gen. Rel. Grav. 48, 131 (2016).

\bibitem{18} B. Pourhassan, M. Faizal, and U. Debnath,   Eur.
Phys. J. C 76, 145 (2016).

\bibitem{19} B. Pourhassan, M. Faizal,   Nucl. Phys. B 913, 834 (2016).

\bibitem{21} \"{O}. \"{O}kc\"{u}, E. Aydiner,   arXiv:1703.09606.

\bibitem{22} S. Upadhyay, B. Pourhassan and H. Farahani,  Phys.
Rev. D 95. 106014 (2017).

\bibitem{23} B. Pourhassan and M. Faizal,   Eur.
Phys. J. C 77, 96 (2017).

\bibitem{24} G. Policastro, D.T. Son  and A.O. Starinets,   JHEP 0209, 043 (2002).

\bibitem{27} P. Kovtun, D.T. Son  and A.O. Starinets,  JHEP 0310,
064 (2003).


\bibitem{29} A. Buchel and J.T. Liu,  Phys. Rev. Lett. 93, 090602 (2004).

\bibitem{new} A. Jawad and M.U. Shahzad,  Eur. Phys. J. C 77, 349 (2017).

\bibitem{30} F. Darabi, F. Felegary and M. R. Setare,   Eur. Phys. J.
C 76, 703 (2016).
\bibitem{32-0}
S. Upadhyay, N. Islam and P. Ganai,  Journal of Holography Applications in Physics 2, 25 (2022).
\bibitem{32} B. Pourhassan, M. Faizal, Z. Zaz  and A. Bhat,  Phys. Lett. B 773, 325
(2017).


\bibitem{Nashed:2020kdb}
G.~G.~L.~Nashed and E.~N.~Saridakis, 
Phys. Rev. D 102,   124072 (2020). 



\bibitem{main} S. H. Hendi, B. Eslam Panah  and S. Panahiyan,  JHEP 1605, 029
(2016).

\bibitem{LIGO2017} B.P. Abbott et al.,   Phys. Rev. Lett. 118, 221101 (2017).

\bibitem{dRGT} C. de Rham, G. Gabadadze  and A.J. Tolley,  Phys. Rev. Lett. 106, 231101 (2011).

\bibitem{deRhamREVIEW2014} C. de Rham,   Living Rev. Rel. 17, 7 (2014).


\bibitem{Cai:2013lqa}
Y.~F.~Cai, F.~Duplessis and E.~N.~Saridakis, 
Phys. Rev. D  90, 064051 (2014)
[arXiv:1307.7150 [hep-th]].

\bibitem{Cai:2014upa}
Y.~F.~Cai and E.~N.~Saridakis, 
Phys. Rev. D 90,  063528 (2014).

\bibitem{Gannouji:2013rwa}
R.~Gannouji, M.~W.~Hossain, M.~Sami and E.~N.~Saridakis, 
Phys. Rev. D  87, 123536 (2013).

{\bibitem{Yan:2018yyz}
W.~Yan, Nonlinearity  {32},  4682  (2019).

\bibitem{Garrione}M. Garrione, arXiv:2306.13788.
}

\bibitem{Cai:2012db}
Y.~F.~Cai, D.~A.~Easson, C.~Gao and E.~N.~Saridakis, 
Phys. Rev. D  87, 064001 (2013).


\bibitem{Vegh2013}D. Vegh, CERN-PH-TH-2012-357 [arXiv:1301.0537].




\bibitem{Cai2015}R.G. Cai, Y.P. Hu, Q.Y. Pan  and Y.L. Zhang, Phys. Rev. D 91, 024032 (2015).

\bibitem{A} S.H. Hendi, S. Panahiyan, R. Mamasani, 
Gen. Rel. Grav. 47, 91 (2015).

\bibitem{B} S. Carlip,   Class. Quant.
Grav. 12, 2853 (1995).

\bibitem{Hawking6} S. W. Hawking, C. J. Hunter and D. N. Page,   Phys. Rev. D 59, 044033 (1999).

\bibitem{P1}
N-ul-islam, P. A. Ganai and S. Upadhyay,  
Prog. Theor. Exp. Phys. 2019, 103B06 (2019).

\bibitem{P2}
Prosenjit Paul, S. I. Kruglov,  Indian J Phys (2023)  [arXiv:2302.05704].




\bibitem{WeinholdI} F. Weinhold, J. Chem. Phys. 63, 2479 (1975).

\bibitem{WeinholdII} F. Weinhold, J. Chem. Phys. 63, 2484 (1975).

\bibitem{RuppeinerI} G. Ruppeiner, Phys. Rev. A 20, 1608 (1979).

\bibitem{RuppeinerII} G. Ruppeiner, Rev. Mod. Phys. 67, 605 (1995).

\bibitem{QuevedoI} H. Quevedo, J. Math. Phys. 48, 013506 (2007).

\bibitem{QuevedoII} H. Quevedo and A. Sanchez, JHEP 09, 034 (2008).

\bibitem{HPEMI} S. H. Hendi, S. Panahiyan, B. Eslam Panah and M. Momennia,
Eur. Phys. J. C 75, 507 (2015).

\bibitem{HPEMII} S. H. Hendi, A. Sheykhi, S. Panahiyan and B. Eslam Panah,
Phys. Rev. D 92, 064028 (2015).

\bibitem{HanC} Y. W. Han and G. Chen, Phys. Lett. B 714, 127 (2012).

\bibitem{BravettiMMA} A. Bravetti, D. Momeni, R. Myrzakulov and A.
Altaibayeva, Adv. High Energy Phys. 2013, 549808 (2013).

\bibitem{Ma} M. S. Ma, Phys. Lett. B 735, 45 (2014).

\bibitem{GarciaMC} M. A. Garc\'{\i}a-Ariza, M. Montesinos and G. F. T. d.
Castillo, Entropy 16, 6515 (2014).

\bibitem{ZhangCY} J. L. Zhang, R. G. Cai and H. Yu, JHEP 02, 143 (2015).

\bibitem{MoLW2016} J. X. Mo, G. Q. Li and Y. C. Wu, JCAP 04, 045 (2016).

\bibitem{BasakCNS} S. Basak, P. Chaturvedi, P. Nandi and G. Sengupta, Phys.
Lett. B 753, 493 (2016).

\bibitem{comp} W. Y. Wen,   Int. J. Mod. Phys. D 26, 1750106 (2017).

\bibitem{dependency} A. Bravetti and F. Nettel,   Phys. Rev. D 90, 044064 (2014).

\bibitem{1201.0463} S. Hollands and R. M. Wald,   Communications in Mathematical Physics 321, 629 (2013).

\bibitem{2007.15401}
A. Chatterjee and A. Ghosh,   Phys. Rev. Lett. 125 (2020) 041302

\bibitem{CANTATA:2021ktz}
E.~N.~Saridakis \textit{et al.} [CANTATA  Collaboration], 
Springer, 2021, ISBN 978-3-030-83715-0
[arXiv:2105.12582 [gr-qc]].

\bibitem{Gonzalez:2011dr}
P.~A.~Gonzalez, E.~N.~Saridakis and Y.~Vasquez, 
JHEP 07, 053 (2012).

\bibitem{Capozziello:2012zj}
S.~Capozziello, P.~A.~Gonzalez, E.~N.~Saridakis and Y.~Vasquez, 
JHEP 02, 039 (2013).





\end{thebibliography}
\end{document}